\documentclass[a4paper, 12pt]{article}  
\usepackage{pdfpages}
\usepackage{lmodern}
\usepackage{graphicx} 
\usepackage{wrapfig} 
\usepackage{textcomp} 
\usepackage[utf8]{inputenc}
\DeclareSymbolFont{letters}{OML}{ztmcm}{m}{it}
\usepackage{color}
\usepackage{bm}
\usepackage{multirow}
\usepackage{listings}
\usepackage{adjustbox}
\usepackage[boxed]{algorithm2e}
\usepackage{subcaption}
\usepackage{multirow} 
\usepackage[OMLmathrm,OMLmathbf]{isomath}
\usepackage{booktabs} 
\usepackage[hang,flushmargin]{footmisc}
\usepackage{pdflscape}
\usepackage{natbib}

\usepackage[T1]{fontenc} 
\usepackage{geometry}
\usepackage{amsmath}
\usepackage{etoolbox} 
\usepackage{amsthm}
\usepackage{amsfonts}
\usepackage{amssymb}
\usepackage[section]{placeins}
\usepackage{diagbox} 
\usepackage{hyperref}
\usepackage{graphicx} 
\usepackage{fixmath} 
\usepackage[scaled=.90]{helvet}
\usepackage{wrapfig}
\usepackage{tikz}

\makeatletter
\renewcommand\@biblabel[1]{\textbf{#1.}} 
\usepackage{enumitem}
\newcommand{\blind}{1}
\usepackage{pifont}
\newcommand{\xmark}{\ding{55}}%
\setlist{nolistsep,leftmargin=*}
\renewcommand{\maketitle}{ 
	\begin{center}
		{\LARGE\@title} 
		
		\vspace{0pt} 
		
		{\large\@author} 
		\\\@date 
		
		\vspace{40pt} 
	\end{center}
}

\begin{document}

	\def\spacingset#1{\renewcommand{\baselinestretch}%
		{#1}\small\normalsize} \spacingset{1}

	
	\if1\blind
	{
	\title{\textbf{Separable and Semiparametric Network-based Counting Processes applied to the International Combat Aircraft Trades}}
\author{Cornelius Fritz$^\ast$, Paul W. Thurner$^\dagger$, Göran Kauermann$^\ast$\hspace{.2cm}\\
	Department of Statistics, LMU Munich$^\ast$\hspace{.2cm}\\ Geschwister Scholl Institute of Political Science, LMU Munich$^\dagger$\hspace{.2cm}}
		\maketitle
	} \fi

	\if0\blind
	{
		\bigskip
		\bigskip
		\bigskip
		\begin{center}
			{\LARGE\bf Dynamic Networks}
		\end{center}
		\medskip
	} \fi
	
	\bigskip

\begin{abstract}
	We propose a novel tie-oriented model for longitudinal event network data. The generating mechanism is assumed to be a multivariate Poisson process that governs the \textsl{onset} and \textsl{repetition} of yearly observed events with two separate intensity functions. We apply the model to a network obtained from the yearly dyadic number of international deliveries of combat aircraft trades between 1950 and 2017. Based on the trade gravity approach, we identify economic and political factors impeding or promoting the number of transfers. Extensive dynamics as well as country heterogeneities require the specification of semiparametric time-varying effects as well as random effects.  Our findings reveal strong heterogeneous as well as time-varying effects of endogenous and exogenous covariates on the \textsl{onset} and \textsl{repetition} of aircraft trade events. 
\end{abstract}
\noindent%
\vfill

\section{Introduction}

Network data capture information on relations between actors. The various types of links between actors in the network encompass stable ties associated with some duration. For example, in political science,  military alliance agreements are active for a certain number of years  \citep{Cranmer2012, Leeds2019}. A different type of link consists of instantaneous bilateral events - like hostile actions measured in real-time \citep{Boschee2015}. Note that instantaneous events can be viewed as the limit case of stable ties if the duration of these ties goes to zero  \citep{buttts2017}. While instantaneous events can happen anytime, they are not always observable in a high resolution of time. Under these circumstances, we can count the instantaneous events occurring in a given time interval, which implies a network-based counting process. We define the respective class of processes as a multivariate counting process that simultaneously guides all dyadic interactions within an event network and dedicate this article to its analysis. Comprehensive monographs and survey articles on statistical network analysis are available by \citet{kolaczyk2009,kolaczyk2017,goldenberg2010, Lusher2012}. Recent overviews of dynamic network modeling can be found in \citet{Fritz2019,kim2018}.

In real-life applications, most networks exhibit dynamics, i.e., structural changes over time are driven by endogenous and exogenous determinants, being covariates that capture the present or past network dependencies and additional information external to the evolution of the network, respectively. One way to conceive the generating process of networks is to represent it as a discrete Markov chain, where the realized path consists of the observed networks, and the state space is the set of all observable networks. The transition probabilities defining the chain are given by a distribution over all possible networks \citep{Robins2001}. For stable ties, this view results in the temporal exponential random graph model (TERGM, \citealp{hanneke2010}).

Alternatively, we can perceive the networks as evolving over time, guided by a continuous Markov process \citep{holland1977}. In this case, network dynamics are often modeled by the stochastic actor-oriented model (SAOM, \citealp{snijders1996}) or in the case of instantaneous events with a precise time stamp by the relational event model (REM) as proposed by \citet{butts2008}. Although modern sensory technology eases the collection of such fine-grained data \citep{lazer2009}, exact continuous information is usually not obtainable for every observed event. In our case, for example, data on the transactions of combat aircraft trades are collected yearly, but the exact time point of each event (e.g., day of delivery) is impossible to verify \citep{SIPRI}. Therefore, instead of observing instantaneous events, we only protocol counts of events during given intervals. Consequently, the resulting event data can be comprehended as valued networks, weighted by the count of events that happened within the given intervals. Though the body of literature on dynamic network models is steadily growing, the consideration of valued dynamic networks is less developed and mainly limited to cross-sectional analyses (see \citealp{desmarais2012,krivitsky2012,wasserman1996,Krivitsky2009}).

In this article, we introduce a tie-oriented model for the analysis of network-based event data. Tie-oriented models assume a bilateral intensity governing the occurrence of events within a dyad, as opposed to actor-oriented models suggested by \citet{stadtfeld2012} and extended in \citet{Hoffman2020,stadtfeld2017}. This approach partitions the intensity into an egocentric sender-specific intensity and a probability selecting the receiver conditional on the sender along the lines of the discrete choice model of \citet{mcfadden1973}. To represent the dynamic evolution of the network-based process, we start with a framework that operates in continuous time at the tie level. Because the ranking of events in our application is not unique due to the lack of exact time stamps, standard REMs \citep{butts2008,vu2011} cannot be readily applied. Therefore, we develop our model under the assumption that the exact ordering of aircraft deliveries within the window of a year is unknown and uninformative. Given that a perennial interplay between policymakers of the involved countries as well as a lengthy order process preludes each trade, this assumption seems reasonable \citep{Snijders2017}. 

Our approach extends existing models in multiple ways. Firstly, we generalize the separable decomposition of network dynamics differentiating between the formation and dissolution of ties introduced by \citet{krivitsky2014,holland1977}. In particular, we extend the separable decomposition to event and count data instead of the continuous specification given in \citet{krivitsky2014,holland1977}, where solely binary and durable ties are regarded. Thereby, we enhance recently introduced \textsl{windowed} effects by \citet{stadtfeld2017}. Furthermore, we propose a semiparametric specification and use penalized B-Splines to obtain flexible time-varying coefficients \citep{eilers1996}. In a similar approach, \citet{Bauer2019} employ non-linear effects to investigate the collaboration between inventors through joint EU patents. \citet{kreiss2017} propose a non-parametric model with time-varying coefficients that necessitates time-continuous observations, although focusing on the estimator's properties as the number of actors goes to infinity. To capture latent actor-specific heterogeneity, we include random effects for each actor in the network differentiating between the sender and receiver of events. As an application case, we focus on the strategically most crucial international deliveries of weapons, namely combat aircraft from 1950 to 2017 \citep{Forsberg1994,sipridata2017}. Combat aircraft comprises all ``unmanned aircraft with a minimum loaded weight of 20 KG'' \citep{siprimeth}.  They are very costly, and the number of units transferred constitutes highly valuable information for military strategists \citep{Forsberg1997}. Therefore, we propose to focus on yearly unit sales as a substantial quantity.

The remainder of this article is structured as follows: the next section formally introduces the tie-oriented model based on a network-based counting process together with extensions to separable, time-varying, and random effects and an estimation procedure. Consecutively,  we introduce the application case and apply our novel method. The paper concludes with Section \ref{sec:conclusion}.

\section{Network-based Counting Process}
\label{sec:Process-based Model}

\subsection{A Framework for Discrete and Continuous Time Event Data}
\label{sec:model_framework}
We start by proposing the model for time-continuous event data, which are observed at discrete time points. We use the temporal indicator $\tilde{t} \in \mathcal{T} = [0, T)$ and mathematically define the network-valued process as a Poisson process on a valued network given by:
\begin{align}
\mathbold{ N } (\tilde{t}) = (N_{ij}(\tilde{t}) \mid i,j \in \lbrace 1, \ldots, n \rbrace),
\label{eq:process}
\end{align}
where $n \in \mathbb{N}$ is the total number of actors in the network. Process \eqref{eq:process} counts the relational events between all actors in the network during the interval $[0,\tilde{t})$. It is characterized by the network-valued intensity rate $\tilde{\mathbold\lambda} (\tilde{t})= (\tilde{\lambda} _{ij}(\tilde{t}) \mid i,j \in \lbrace 1, \ldots, n \rbrace)$. The $(i,j)$th entry of this intensity is defined as the probability that we observe an instantaneous jump of size 1 in $N_{ij}(\tilde{t})$. Heuristically, this is the probability of the occurrence of a directed event from actor $i$ to $j$ at time point $\tilde{t}$. By definition we set $\tilde \lambda_{ii}(\tilde{t}) = 0 ~ \forall~ i \in \lbrace 1, \ldots, n \rbrace$ and $\tilde{t} \in \mathcal{T}$. 

Assuming the process is observed at discrete time points $t \in \lbrace 0, \ldots, T \rbrace$ leads to the time-discrete observations $\mathbold Y_{t}$, which are defined as cumulated events through:
\begin{align*}
\mathbold Y_t = \mathbold N(t)-\mathbold N(t-1)~ \forall~  t \in \lbrace 1, \ldots, T \rbrace, 
\end{align*}
with $\mathbold{N} (0)$ set to 0. Based on the properties of a Poisson process, these increments follow a matrix-valued Poisson-distribution: 
\begin{align}
\mathbold Y_t \sim \text{Pois}\Bigg(\int_{t-1}^t\tilde{\mathbold\lambda} (\tilde u) d \tilde u\Bigg) ~ \forall~  t \in \lbrace 1, \ldots, T \rbrace.
\label{eq:poisson-process}
\end{align}
Given that the exact orderings of events within each observation window are not known and assumed to be uninformative, 
the integrated intensity on the time interval $(t-1,t]$ simplifies to a constant, so that $\int_{t-1}^t\tilde{\mathbold\lambda} (\tilde u) d \tilde u = \mathbold\lambda(t)$ holds. 
Accordingly, we define the observed values of $\mathbold Y_t$ as $\mathbold y_t$.  As a result of \eqref{eq:poisson-process}, the waiting times between subsequent events follow an exponential distribution. Therefore, our model is equivalent to the REM as introduced in \citet{butts2008} in the special case where $\parallel \mathbold y_t \parallel_1 = 1 ~ \forall~  t \in \lbrace 1, \ldots, T \rbrace$ holds under piece-wise constant intensities.  

Generally, we are interested in modeling $\mathbold\lambda(t)$ conditional on the past network topology and exogenous covariates, which are denoted by $x_t$.  Covariates can be node-specific (regarding either a feature of the sender or receiver), dyadic (regarding a relation between the sender and receiver), or global (regarding the complete network). Building on a first-order Markov property, we allow the intensity to depend on the past network behavior and exogenous covariates through: 
\begin{align}
Y_{ij,t} \sim \text{Pois}\big(\lambda_{ij}(t ,\mathbold y_{t-1}, \mathbold x_{t-1})\big)~ \forall~  t \in \lbrace 1, \ldots, T \rbrace;~  i,j \in \lbrace 1, \ldots, n \rbrace, i \neq j.
\label{eq:poisson-process-tie}
\end{align}
This is equivalent to the assumption of dyadic independence of events to occur in each time interval given information on the past and exogenous covariates. Similar assumptions were made by \citet{lebacher2019_sep} in the context of separable TERGMs (STERGM, \citealp{krivitsky2014}). \citet{almquist2014} justify this method for network panel data where little simultaneous dependence between possible ties is present. For our application to the international combat aircraft trades, this can be legitimized by the long time span of aircraft trades between the order and delivery of units\footnote{We further provide a descriptive analysis in the Supplementary Material to demonstrate high positive auto-correlation of the endogenous covariates between consecutive years, therefore they are a reliable proxy of simultaneous dependence.}.

Accordingly, we specify the intensity in time-varying semiparametric form through: 
\begin{align}
\lambda_{ij}(t, \mathbold y_{t-1}, \mathbold x_{t-1} ) = \lambda_0(t) \text{exp}\lbrace \theta(t)^\top s_{ij}(\mathbold y_{t-1}, \mathbold x_{t-1})\rbrace ~  \forall~  t \in \lbrace 1 , \ldots, T \rbrace, \label{eq:intensity_final}
\end{align}
where $\lambda_0(t)$ is the baseline intensity, $s_{ij}(\mathbold y_{t-1}, \mathbold x_{t-1})$ is a multidimensional vector consisting of network statistics and theoretically derived exogenous covariates in $t-1$. We discuss different specifications of the statistics in Section \ref{sec:application} where we describe the application case in more detail. The coefficient vector $\theta(t)$ is possibly time-varying and needs to be estimated from the data. 

In many application cases, compositional changes of the actor set occur. To compensate for this phenomenon in the model, we include indicator functions similar to risk indicators in time-to-event analysis \citep{Kalbfleisch2002}. To be specific, we multiply the intensity by an indicator function, determining whether actors $i$ and $j$ are both present in the network at time $t$: 
\begin{align}
\lambda_{ij}(t, y_{t-1}, x_{t-1} ) = \mathbb{I}(i,j \in \mathcal{R}_t) \lambda_0(t) \text{exp}\lbrace \theta(t)^\top s_{ij}(y_{t-1}, x_{t-1})\rbrace ~  \forall~  t \in \lbrace 1 , \ldots, T \rbrace, 
\label{eq:intensity_final2}
\end{align}
with $\mathcal{R}_t$ denoting the set of actors partaking in the network at time point $t$. By including the indicator functions $\mathbb{I}(i,j \in \mathcal{R}_t)$, we decompose our observed network into a stochastic and deterministic component. The latter component consists of structural zeros at time point $t$ in the modeled network between all actors where at least one side is not present. With these actor set changes, the possible range of the network statistics changes as well, leading to values which are not scaled coherently for a comparison across years. To solve this issue, we divide all network statistics by their maximal value to allow for a cohesive interpretation. 



\subsection{Extensions}
\label{sec:Extensions}

\subsubsection{Separability Assumption}
\label{sec:Separability Assumption}
Interaction patterns are commonly substantially different for already linked and still unlinked actors. To adequately capture this characteristic, \citet{holland1977} proposed a process-based model for binary ties taking the values ``0'' or ``1'' by two separate intensity functions. One intensity toggles entries from ``0'' to ``1'' (formation of ties) and another one from ``1'' to ``0'' (dissolution of ties). 
Thereby, separate and potentially differential effects of statistics depending on previous interaction behavior are enabled. This model, henceforth called \textsl{separable} model, was later adopted to the SAOM by incorporating a \textsl{so-called} gratification function \citep{Snijders1997,Snijders2003} and to the TERGM by extending it to the separable TERGM \citep{krivitsky2014}. However, one should keep these \textsl {separable} models apart from the \textsl {separability condition} introduced in \citet{almquist2014}. In the following, we combine the framework of relational event models with the separability approach as coined by \citet{krivitsky2014}.  

More specifically, we postulate two different conditions for the network-based process under which the effect of all covariates changes. One condition governs events between unlinked actors and is characterized by the \textsl{onset} intensity. The second condition only regards events among actors that already interacted with each other and is driven by the \textsl{repetition} intensity. In accordance with the the Markov assumption specified in \eqref{eq:intensity_final}, we define the \textsl{onset} intensity at time $t$ to control all events which did not occur in $y_{t-1}$. Accordingly, the \textsl{repetition} intensity drives the events that did occur at least once in $y_{t-1}$. This can be incorporated by splitting the intensity into two conditional intensities: 
\begin{align}
\lambda_{ij}(t,\mathbold y_{t-1}, \mathbold x_{t-1}) = \begin{cases}\lambda_{ij}^+(t,\mathbold  y_{t-1}, \mathbold x_{t-1}), &\text{if $y_{ij,t-1} = 0$}\\ \lambda_{ij}^-(t, \mathbold y_{t-1}, \mathbold x_{t-1}), &\text{if $y_{ij,t-1} >0 $}
\end{cases},
\label{eq:REMseparable}
\end{align}
where $\lambda_{ij}^+(t, \mathbold y_{t-1}, \mathbold x_{t-1})$ and $\lambda_{ij}^-(t, \mathbold y_{t-1}, \mathbold x_{t-1})$ are defined along the lines of \eqref{eq:intensity_final} and specified by the corresponding time-varying parametric effects $\theta^+(t)$ and $\theta^-(t)$ jointly represented by $\theta(t) = \big(\theta^+(t), \theta^-(t)\big)$. The possibly overlapping vectors of statistics are denoted accordingly as $s_{ij}^+( \mathbold y_{t-1}, \mathbold x_{t-1})$ and $s_{ij}^-( \mathbold y_{t-1}, \mathbold x_{t-1})$, respectively. Setting $s_{ij,0}^+(\mathbold y_{t-1},\mathbold  x_{t-1})= 1$ enables the inclusion of a time-varying intercept $\lambda_0^+(t) = \exp \lbrace \theta_0^+(t) \rbrace$ in the \textsl{onset} model, this holds similarly for the \textsl{repetition} model. Consecutively, the complete separable model is given by replacing \eqref{eq:intensity_final} with 
\begin{align}
\lambda_{ij}(t,\mathbold y_{t-1}, \mathbold x_{t-1}) =& \exp \Big\lbrace \mathbb{I}(y_{ij,t} = 0) \big[\theta^+(t)^\top s_{ij}^{+}(y_{t-1}, x_{t-1})\big] \nonumber\\
& \hspace*{1cm} + \mathbb{I}(y_{ij,t} > 0)  \big[\theta^-(t)^\top s_{ij}^-(y_{t-1}, x_{t-1}) \big] \Big\rbrace \nonumber\\
=& \exp \big\lbrace \theta(t)^\top s_{ij}(y_{t-1}, x_{t-1}) \big\rbrace 
\label{eq:REMseparable_fin}
\end{align}
where $\theta(t)=\big(\theta^+(t), \theta^-(t)\big)$ and 
\begin{align*}
s_{ij}(y_{t-1}, x_{t-1}) = \big(\mathbb{I}(y_{ij,t} = 0) \cdot s_{ij}^{+}(y_{t-1}, x_{t-1}), \mathbb{I}(y_{ij,t} > 0) \cdot s_{ij}^{-}(y_{t-1}, x_{t-1}) \big).
\end{align*}

\subsubsection{Spline-based Time-Varying Effects}

\label{sec:Time-Varying Effects}


Let the $k$th component of statistic $s_{ij} (\mathbold y_{t-1}, \mathbold x_{t-1})$ be defined as $s_{ij,k}(\mathbold y_{t-1}, \mathbold x_{t-1})$ with the matching coefficient $\theta_k(t)$.  We expand each component $\theta_k(t)$ in a semiparametric way by replacing it with a B-Spline basis function (see \citealp{DeBoor2001}). More specifically, we place equidistant knots on a grid in $\mathcal{T}$, where the number of knots can chosen relatively high \citep{Kauermann2011}. In principle, we could choose individual grids for each component of $\theta(t)$, but for the sake of a simple notation, we select the same one for all covariates. We now rewrite each coefficient as:
\begin{align}
\theta_k(t) =B(t) \alpha_k ~ \forall ~ k \in \lbrace 0, \ldots, K \rbrace,
\label{eq:p_spline}
\end{align}
where $B(t) \in \mathbb{R}^q$ is the B-spline basis evaluated at $t$ and $\alpha_k \in \mathbb{R}^q$ denotes the corresponding coefficient vector. In our context, $q$ constitutes the dimension of the B-spline basis and hence gives the number of separate B-spline bases used for each covariate.  To obtain a smooth fit we penalize the difference of adjacent basis coefficients $\alpha_k$ as proposed by \citet{eilers1996}. This leads to the overall penalized log-likelihood function: 
\begin{align}
\ell_p(\alpha_0, \ldots, \alpha_K,\gamma_0, \ldots, \gamma_K) \propto \sum_{t = 1}^T \sum_{i,j \in \mathcal{R}_t} \big( y_{ij,t} \log (\lambda_{ij,t}) - \lambda_{ij,t}\big) - \frac{1}{2} \sum_{k = 0}^K \gamma_k \alpha_k^\top D_k \alpha_k, \label{eq:likelihood_est}
\end{align}
with $\lambda_{ij,t}= \lambda_{ij}(t,\mathbold y_{t-1}, \mathbold x_{t-1})$. The penalty results from the quadratic form with penalty matrix $D_k$ constructed from pairwise differences of the spline coefficients and $\gamma_k$ as the penalty (and hence tuning) parameter. This vector $\gamma = (\gamma_1, \ldots, \gamma_K)$ controls the smoothness of the fit and is chosen data based following a mixed model approach as described in detail in \citet{Ruppert2003a}, see also \citet{wood2017}. The incorporation of a penalization in \eqref{eq:likelihood_est} results in a biased estimator and a \textsl{so-called}  bias-variance tradeoff, which is thoroughly discussed for penalized spline smoothing in \citet{Ruppert2003a}.  \citet{Kauermann2011} extend the theoretical results towards a data-driven finite-sample version and \citet{Kauermann2009} show that the estimates from \eqref{eq:likelihood_est} are consistent. 


\subsubsection{Accounting for Nodal Heterogeneity}
\label{sec:Accounting for Nodal Heterogeneity}
The specification of the model introduced so far implicitly implies that the nodal heterogeneity is fully captured by the structural statistics $ s_{ij}(\mathbold y_{t-1}, \mathbold x_{t-1})$. As already thoroughly discussed by \citet{thiemichen2016} or \citet{steffensmeier2018}, this can be considered a questionable assumption. It seems, therefore, advisable to include sender- and receiver-specific random effects to account for unobserved heterogeneity. Let therefore $u_i^{S}$ denote a latent sender-specific effect of actor $i$ and $u_j^{R}$ the receiver-specific effect of actor $j$. This leads to the heterogeneous intensity
\begin{align}
\lambda_{ij}(t, \mathbold y_{t-1},\mathbold x_{t-1},u^S,u^R) = \lambda_{ij}(t,\mathbold y_{t-1}, \mathbold x_{t-1}) \text{exp}\lbrace u_i^{S} +u_j^{R}\rbrace ~ \label{eq:random}\forall~  t \in \lbrace 1, \ldots, T \rbrace. 
\end{align}
We assume $u^{S} = (u_1^{S}, \ldots, u_n^{S})^\top \sim N(0, I_n \tau^2_S)$ and $u^{R} = (u_1^{R}, \ldots, u_n^{R})^\top \sim N(0,  I_n\tau^2_R)$ with $I_n$ as the $n \times n$ identity matrix. The expression $\lambda_{ij}(t,\mathbold y_{t-1}, \mathbold x_{t-1})$ may be specified through \eqref{eq:intensity_final} or \eqref{eq:REMseparable_fin}. Conditional on the random effects $u^S$ and $u^R$, the distributional assumption \eqref{eq:poisson-process-tie} still holds: 
\begin{align}
&Y_{ij}(t)\mid u^S,u^R \sim \text{Pois}\big(\lambda_{ij}(t, \mathbold y_{t-1},\mathbold x_{t-1},u^S,u^R)\big)\label{eq:assumption_random}  \\
&\forall~  t \in \lbrace 1, \ldots, T \rbrace;~  i,j \in \lbrace 1, \ldots, n \rbrace, i \neq j \nonumber,
\end{align}
where $\lambda_{ij}\big(t, \mathbold y_{t-1}, \mathbold x_{t-1},u^S,u^R \big)$ is specified in \eqref{eq:random}. 

\subsection{Estimation}
\label{sec:estimation}

The vector-valued function $\theta(t) = \big(\theta^+(t),\theta^-(t) \big)$ is estimated by finding the argument maximizing the penalized likelihood resulting from \eqref{eq:assumption_random} and viewing the penalty on coefficient vector $\alpha$ as a improper prior distribution. This leads to a generalized additive mixed model, which is extensively discussed in \citet{wood2017,Ruppert2003a, ruppert2009}.
To leverage the advanced optimization techniques proposed for this model class, we initially calculate all covariates  $s_{ij}(\mathbold y_{t-1}, \mathbold x_{t-1})$ for each actor-tuple and at each point in time. By doing that, we transform the data into a generalized version of the \textsl{so-called} counting-process representation, which is known from time-to-event analysis \citep{Tutz2016,friedman1982,whiteheadd1980}. For each snapshot of the event network at time point $t$, this procedure generates a design matrix of $|\mathcal{R}_t|$ conditionally independent observations with a target variable $y_{ij,t}$ expressing the number of events that occurred between a specific tuple of actors and covariates given by $s_{ij}(\mathbold y_{t-1}, \mathbold x_{t-1})$. 

For the estimation, we use the versatile $\mathtt{R}$ package $\mathtt{mgcv}$ (\citealp{wood2017}, version 1.8-31). Thereby, we follow \citet{wood2017b} who enhance the pseudo-quasi-likelihood (PQL) method by \citet{breslow1993} for the analysis of larger data sets. The main extensions are threefold:
\begin{enumerate}
	\item  The tuning parameters $\gamma$ are not estimated until convergence in each iteration of the estimation procedure but updated by only one Newton step.
	\item Efficient methods for computing the matrix cross-products in each iteration are run in parallel \citep{Li2019}.
	\item The covariates are discretized along a marginal grid. Hence, the design matrices for the smooth covariates take significantly less memory. 
\end{enumerate}
\citet{wood2017b} describe the method in detail as it is implemented in the function  $\mathtt{bam}$ of the already mentioned  $\mathtt{R}$ package. Well-calibrated frequentist confidence bands for the estimated function $\theta(t)$ are guaranteed by Bayesian large sample properties \citep{Wood2013}.  

\section{Application}
\label{sec:application}
\FloatBarrier

\subsection{Data }
\label{sec:Data Description}
\begin{figure}[t!]
	\FloatBarrier
	\centering
	\captionsetup[subfigure]{labelformat=empty}
	\begin{subfigure}[c]{0.47\textwidth}
		\center
		\includegraphics[width=\linewidth]{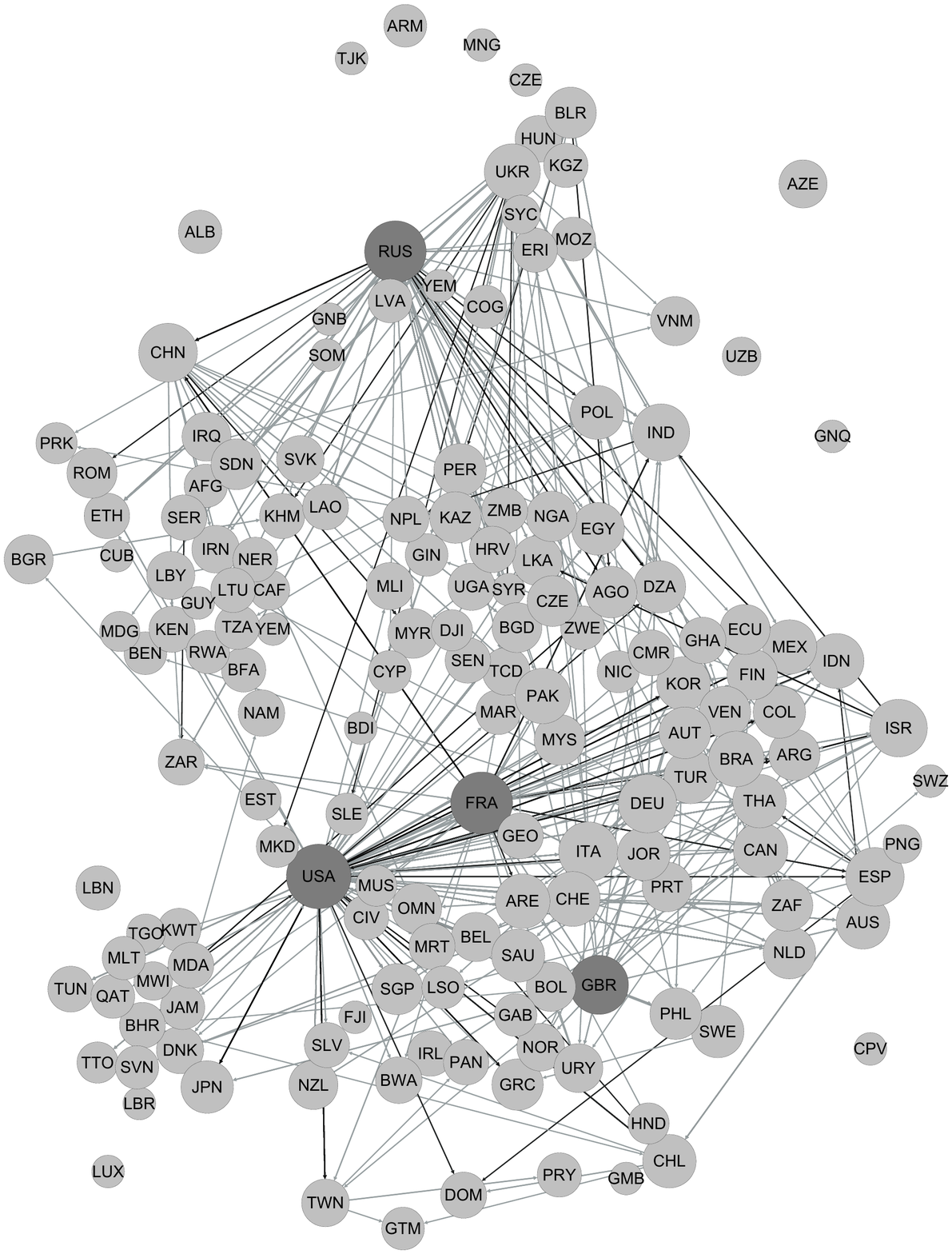}
		\subcaption{Years 1995 - 2000}
		\FloatBarrier
	\end{subfigure}
	\begin{subfigure}[c]{0.47\textwidth}
		\center
		\includegraphics[width=\linewidth]{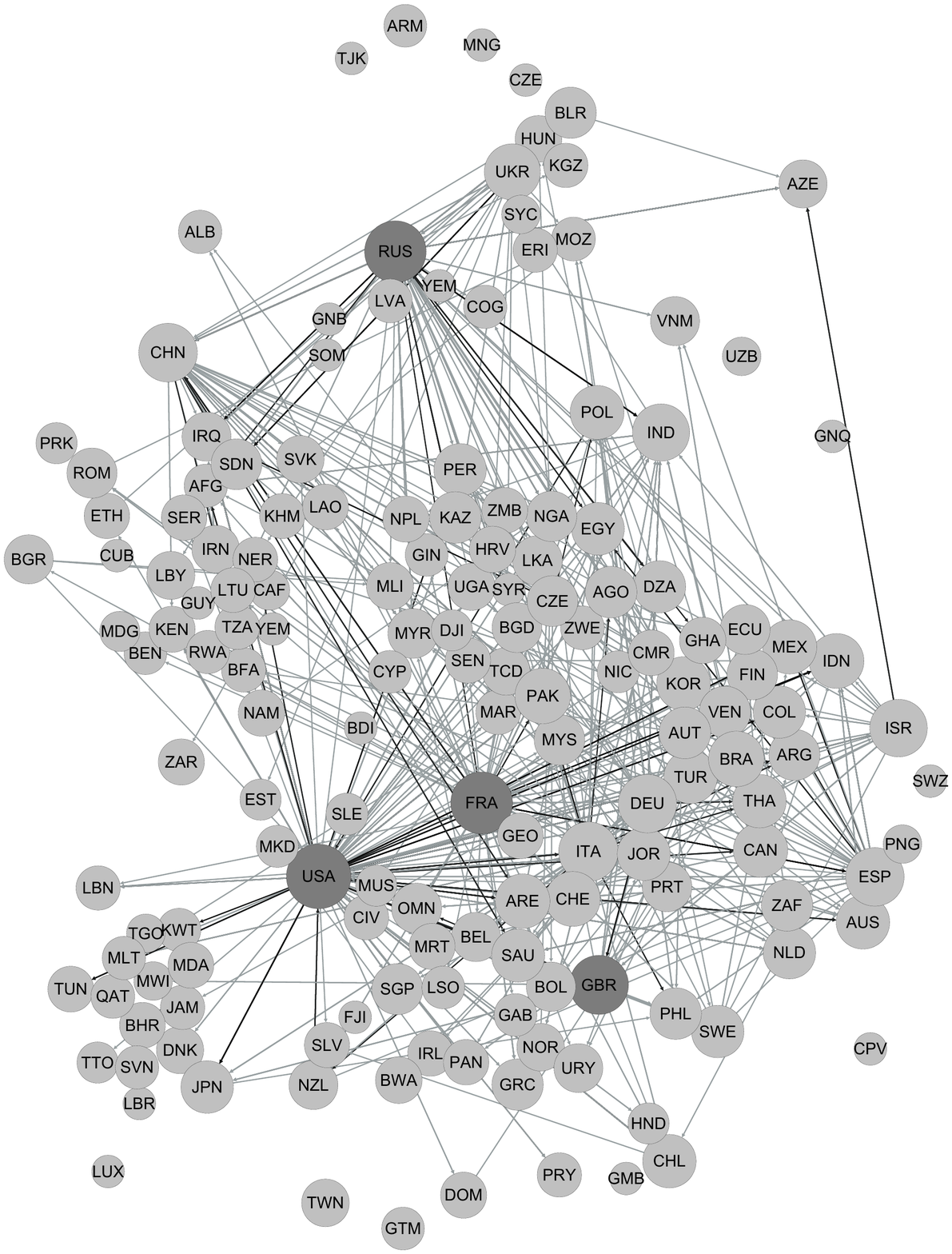}
		\subcaption{Years 2012 - 2017}
	\end{subfigure}
	\caption{The international network of combat aircraft trades in two periods. Node size is proportional to the sum of involved deals and the grey-scale of each tie indicates the aggregated amount of deals in the specific time frame. The
		labels of the nodes are the ISO3 codes of the respective countries. The four major sender countries are drawn in a darker shade.}
	\label{fig:1951}
\end{figure}

So far, quantitative work on the international arms trade utilizing statistical network analysis has mostly been restricted to binarized networks.  Here, the occurrence of a trade relationship between two countries in a specific year was modeled conditional on endogenous and exogenous statistics by the gravity model of trade by employing TERGMs and extensions of it \citep{lebacher2019_sep,Thurner2019}. Contrary,  \citet{Lebacher2019} fit a network disturbance model on the yearly aggregated trend indicator values (TIV, \citealp{siprimeth}) of the international arms trades, maintaining the valued character of deliveries. All these contributions rely on data provided by the Stockholm International Peace Research Institute (\citealp{sipridata2017}), and they consider each type of major conventional weapons indiscriminately. 

In the following, we concentrate on the counts of combat aircraft deliveries, as reported in the SIPRI data, where each combat aircraft delivery is perceived as an event. We focus on the transfers of aircraft because these weapon systems usually incorporate the highest technological sophistication. Therefore, they are being restricted to close allies. Furthermore, they are of crucial strategic importance for international deterrence and counterinsurgency in intrastate conflict \citep{Hoeffler2016,Mehrl2020}. Lastly, their sizes and cost make the available data highly reliable \citep{Forsberg1994,Forsberg1997}. Previous research on combat aircraft trade was limited to the quantitative analysis of a small subset of countries or fighter programs \citep{Hoeffler2016,Vucetic2011,Vucetic}. Contrasting these endeavors, we take a global point of view on the combat aircraft trade. Here, a closer look at the data reveals how countries commonly partition major deals with their stable trade partners into multiple deliveries occurring over the span of several years. For instance, the United States and Japan  signed a deal in 1984 comprising 32 quantities of aircraft, which were realized between 1988 and 2016. The additional information provided by this segmentation of trade deals into isolated deliveries would be lost when only regarding binarized networks\footnote{In the Supplementary Material, we deliver the results regarding alternative models for the data. Overall, there is no relevant difference to the findings presented subsequently.}.


\nocite{csardi2006,ICWSM09154}

\begin{figure}[t!]
	\FloatBarrier
	\centering
	\includegraphics[width=\linewidth]{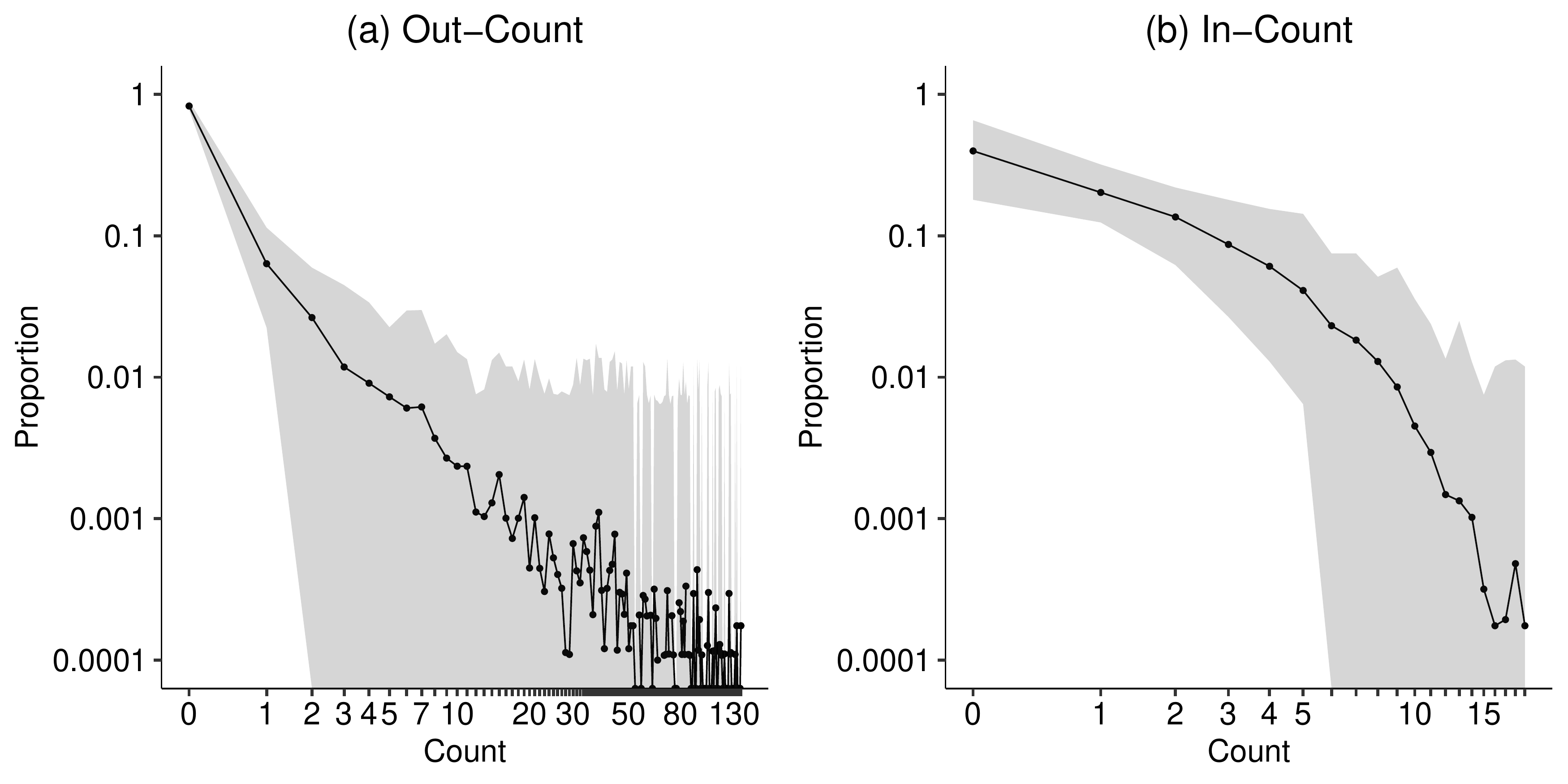}
	\caption{Distributions of the Out- and In-Counts for all included countries concatenated over all years. The shaded area represents the minimum and maximum of the observed values. Both graphs are represented on a logarithmic scale.}
	\label{fig:outdeg2}
\end{figure}
\FloatBarrier
Two examples of the network representing aggregated events over six years are depicted in Figure \ref{fig:1951}. Generally, the networks exhibit a structure with hubs around the United States (USA), Russia (RUS), France (FRA), and United Kingdom (UK). Coincidentally, this set of countries also demonstrate the highest average hub-scores over time \citep{Kleinberg1999}. Analog to the distribution of the in- and out-degrees in binary networks, we can examine the distribution of the concatenated in- and outgoing event counts for all years. We call the respective statistics in- and out-count, although they are equivalent to the generalized degree proposed by \citet{Opsahl2010}. The empirical distribution of those statistics enables a better understanding of the topology of the observed networks. Figure \ref{fig:outdeg2} (a) suggests a strong centralization in the outward event count distribution. Some countries are the sender of up to 130  deliveries in one year. Still, on average, $82\%$ of the countries do not export. The inward count distribution is not as skewed and centralized, as shown in Figure \ref{fig:outdeg2} (b). Nonetheless, the mode of the distribution is still at zero. 


\FloatBarrier
\subsection{Model Specification}
\begin{figure}[!]
	\FloatBarrier
	\centering
	\begin{tabular}{cccc}
		\hspace{-0.5cm} \includegraphics[width=0.24\linewidth, page = 6]{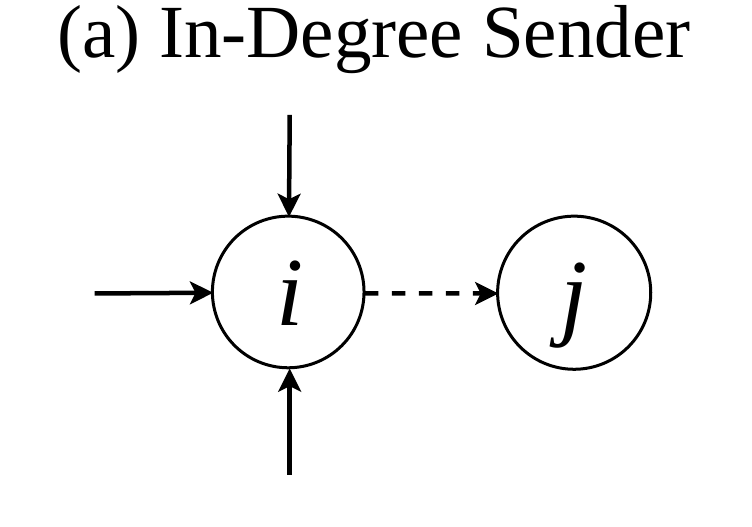} 	&\includegraphics[width=0.24\linewidth, page = 8]{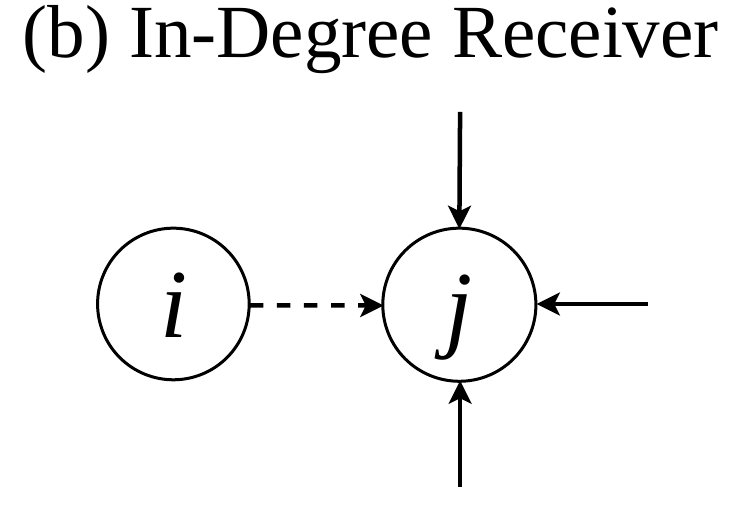}   & \includegraphics[width=0.24\linewidth, page = 5]{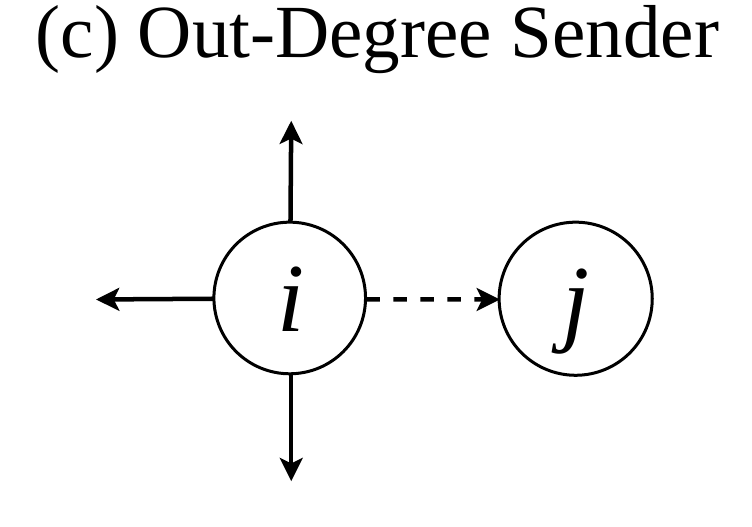} & \includegraphics[width=0.24\linewidth, page = 7]{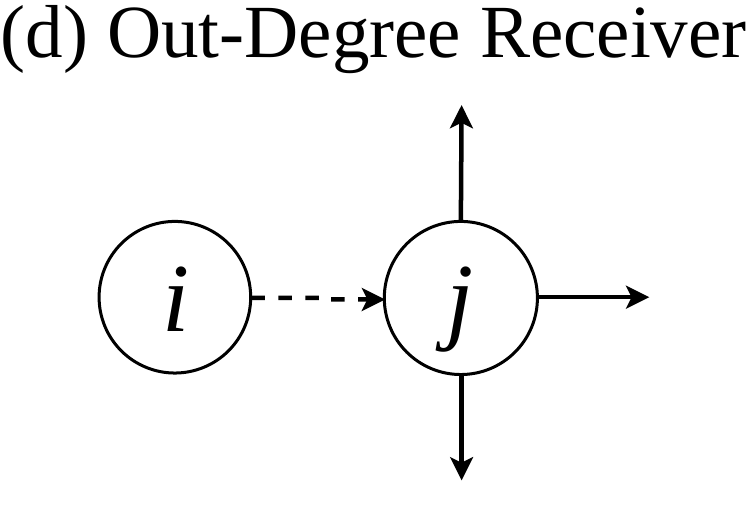} \\
		\multicolumn{4}{c}{ \begin{tabular}{ccc}
				\includegraphics[width=0.24\linewidth, page = 4]{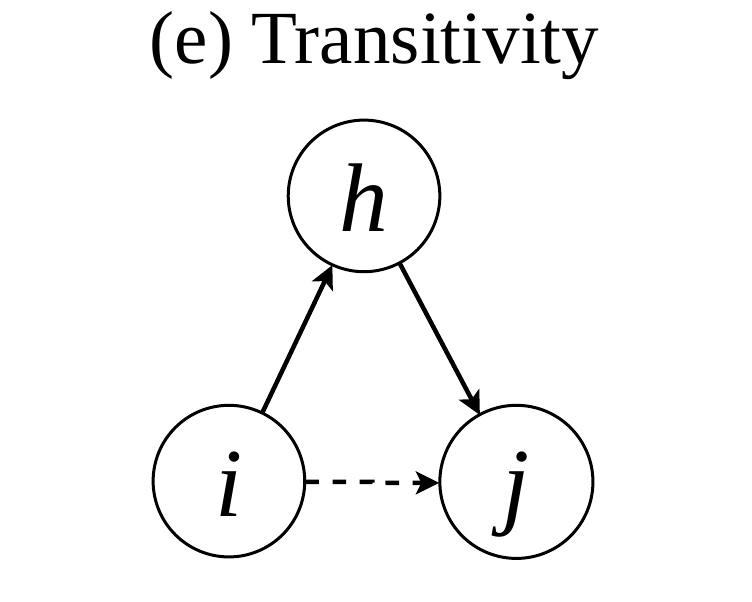} \hspace*{-0.6cm}	&\includegraphics[width=0.24\linewidth, page = 5]{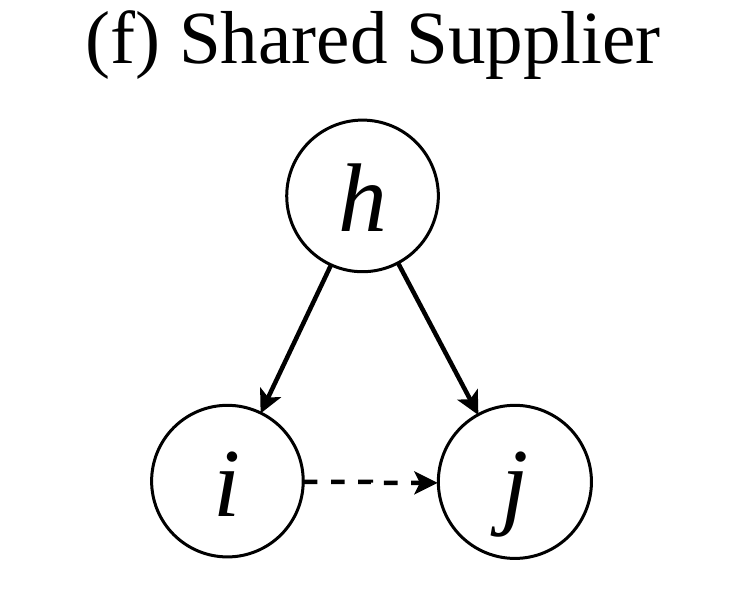}  & \includegraphics[width=0.24\linewidth, page = 6]{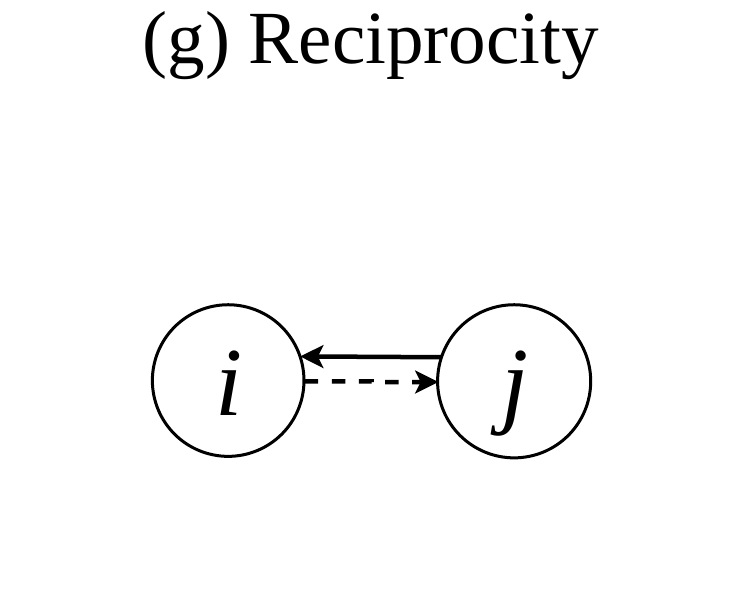} \vspace{-0.2cm}
		\end{tabular}  } 
	\end{tabular} 
	\caption{Graphs consisting of three arbitrary actors $i$,$j$, and $h$ that illustrate the included triangular and dyadic covariates in the first row. Dashed arrows represent the event that is modeled and solid arrows in $t-1$.  }
	\label{fig:newstatistics_centr}
\end{figure}

\FloatBarrier
We now employ the outlined model to the international combat aircraft trade network spanning from 1950 to 2017. The event networks are observed yearly. In this context, $y_{ij,t}$ denotes the number of observed combat aircraft units delivered in year $t$ between country $i$ and $j$ and its distribution follows from \eqref{eq:poisson-process-tie}. Given this information, we estimate the time-continuous intensities of all country-dyads, which are per assumption governed by the \textsl{repetition} intensity if the respective countries traded in the previous year and by the \textsl{onset} intensity otherwise as defined in \eqref{eq:REMseparable_fin}\footnote{As a robustness check, we compare different time frames to define which events are driven by the \textsl{onset} and \textsl{repetition} intensity, e.g., having delivered combat aircraft the last one or two years in the Supplementary Material.}. All network actors are countries, and an event represents the delivery of combat aircraft between them.  To appropriately capture interdependencies of the observed event counts, we incorporate a wide range of endogenous statistics, whose mathematical representation is given in Table \ref{tbl:endo_variables} and visualized in Figure \ref{fig:newstatistics_centr}. Generally, we define all non-binary structural statistics to be bounded between 0 and 100 to guarantee a consistent interpretation independent of the varying network size and prevent the implied autoregressive counting process from unrealistic behavior \citep{Gjessing2010}.   

\begin{table}[!]
	\center
	\caption{Mathematical formulations of the structural covariates as calculated for $s_{ij}(\mathbold y_{t-1}, \mathbold x_{t-1})$.  The number of countries that are present in the network at time point $t$ is denoted by $n_t$. The identifying letters concern the respective graphical illustrations in Figure \ref{fig:newstatistics_centr}.}
	\label{tbl:endo_variables}

	\begin{tabular}{l  c}
		Name & Mathematical Representation  \\ \hline
		(a) In-Degree Sender &$  \frac{100}{n_t -1}\sum_{h = 1} ^n \mathbb{I}(y_{hi,t-1} > 0)$  \\
		(b) In-Degree Receiver &$  \frac{100}{n_t -1} \sum_{h = 1} ^n \mathbb{I}(y_{hj,t-1} > 0)$  \\
		(c) Out-Degree Sender  & 	$ \frac{100}{n_t -1}\sum_{h = 1} ^n \mathbb{I}(y_{ih,t-1} > 0)$ \\
		(d) Out-Degree Receiver & $  \frac{100}{n_t -1} \sum_{h = 1} ^n \mathbb{I}(y_{jh,t-1} > 0)$\\
		(e) Transitivity & $  \frac{100}{n_t - 2} \sum_{h = 1} ^n \mathbb{I}(y_{ih,t-1} > 0)  \mathbb{I}(y_{hj,t-1} > 0)$ \\
		(f) Shared Supplier & $  \frac{100}{n_t-2} \sum_{h = 1} ^n \mathbb{I}(y_{hi,t-1} > 0)  \mathbb{I}(y_{hj,t-1} > 0)$ \\
		(g) Reciprocity & $\mathbb{I}(y_{ji,t-1} > 0)$\\
	\end{tabular} 
	
\end{table}


As already investigated in multiple applications \citep{Snijders2003,Newman2002}, the degree structure plays a crucial role in the observed event network. In the case of directed events, the in- and out-degree of a country determine its relative location in the network  \citep{Wasserman1994}. In our application, the degrees reflect the number of different countries with whom a specific country had at least one transaction in a particular year as an importer (in-degree) and exporter (out-degree). To reveal the impact of these measures on the intensity of observing an event, we include four degree-related statistics concerning the sender and receiver in our specification, as illustrated in Figure \ref{fig:newstatistics_centr} (a) - (d). For instance, one can interpret a positive effect of the sender's out-degree as the tendency to trade with countries that are already sending a lot in the previous year. 

Besides degree-based statistics, \citet{Holland1971,Davis1970} highlight the role of triangular structures in networks. When adapted to event relations, it refers to the change in intensity of an event between countries $i$ and $j$, if they are indirectly connected by an additional two-path, i.e., third country. Since the aircraft deliveries between countries are directed, there are multiple ways to define two-paths. We incorporate two triadic structures: {transitivity}, Figure \ref{fig:newstatistics_centr} (e), and {shared supplier}, Figure \ref{fig:newstatistics_centr} (f). While {transitivity} in an event network suggests that already having observed a delivery from country $i$ to $k$ and $k$ to $j$ affects the intensity of an event from $i$ to $j$, the {shared supplier} mechanism reflects the tendency towards trading with countries that import combat aircraft from a common exporter. These triangular structures were the only variants found to be relevant for the trade of combat aircraft. Likewise, we control for reciprocity, which is the tendency of countries to respond to previous events directed at them, Figure \ref{fig:newstatistics_centr} (g). 

Political economy models of arms trade \citep{Levine1994,Thurner2019} as well as the gravity model of arms trade guide the selection of appropriate exogenous covariates. \citet{Thurner2019,Akerman2014} included the dyadic distance in kilometers between the capitals of country $i$ and $j$ as well as the logarithmic gross domestic product (GDP in US $\$$) of the sender and receiver countries as covariates in the model. \citet{Pamp2018,lebacher2019_sep}  emphasize the impact of military expenditures as a proxy for the Newtonian power of attraction, which we include in logarithmic form as a sender- and receiver-specific covariate. The respective yearly data was collected by \citet{SIPRI} in US $\$$ and combined by \citet{Nordhaus2012} with data from \citet{singer1972}. We use this combined data set, but employ linear interpolation if at least 60$\%$ of the time series for a specific country is observed. Moreover, we incorporate two dyadic variables controlling whether country $i$ and $j$ signed an alliance treaty or are similar to each other in terms of their regimes in power, following \citet{johannsen2014,Thurner2019}. The alliance treaty obligations and provisions project identified military alliance agreements \citep{Leeds2019} and we  operationalize regime dissimilarity by the absolute difference in the Polity IV scores of countries $i$ and $j$ \citep{marshall2017}. This measure indicates all countries' year-wise regime characteristics and takes values from -10 (strongly autocratic) to 10 (strongly democratic). Thus, the absolute differences lie between 0 (strong similarity) and 20 (strong dissimilarity) for each country-dyad and year. The sources and used period of all incorporated exogenous covariates are described in more detail in the Supplementary Material. 

\FloatBarrier
\subsection{Results}\label{sec:res}
\FloatBarrier
\subsubsection{Fixed Effects}\label{sec:fixed_eff}
\begin{figure}[!]\centering
	\FloatBarrier
	\includegraphics[trim={0cm 0cm 0cm 0cm},clip,width=\textwidth]{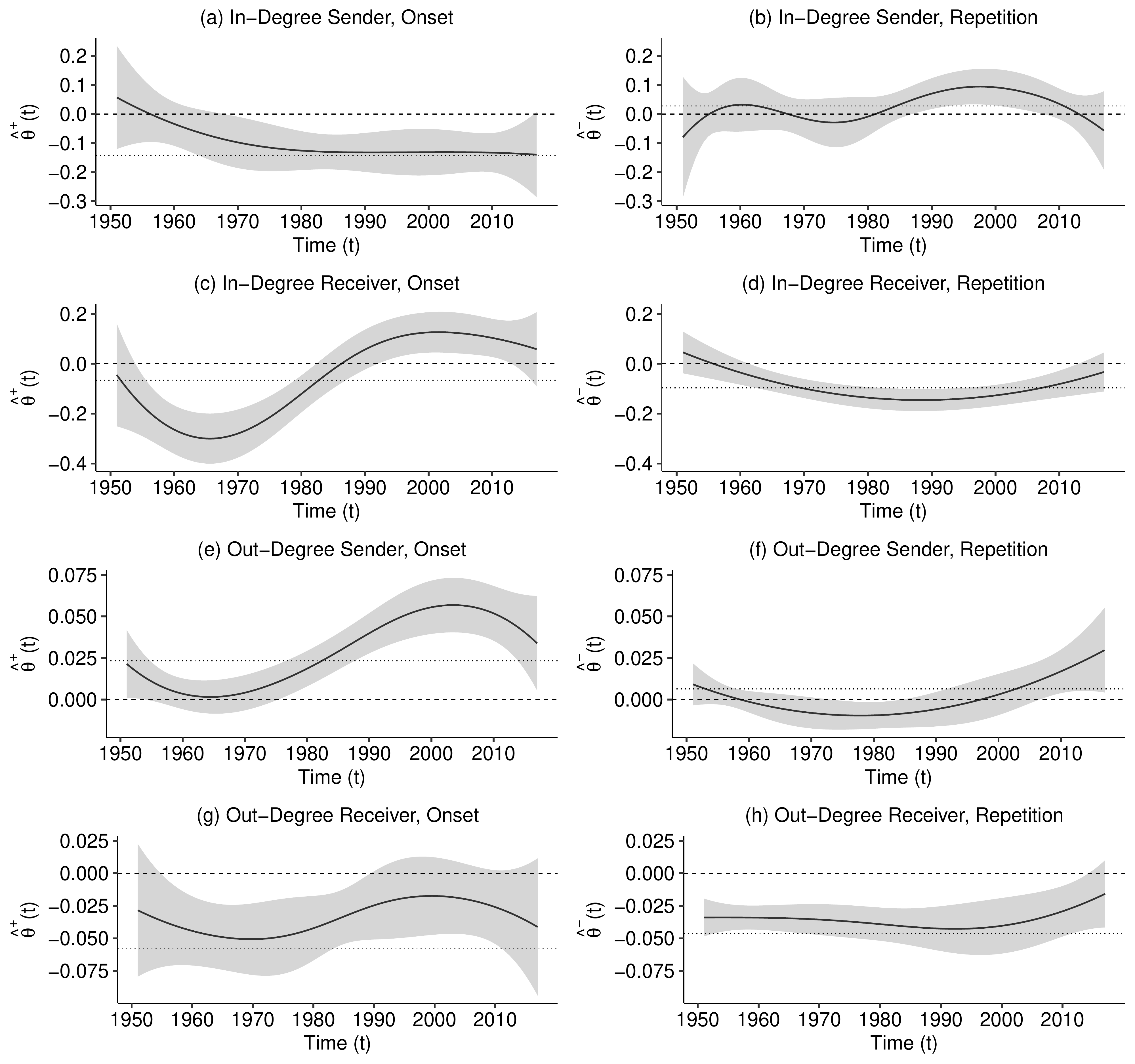}
	\caption{Results of endogenous statistics relating to centrality. The shaded area indicates the $95\%$ confidence bands of the estimates and the dotted horizontal lines represent the time-constant parameters.}
	\label{fig:end1}
\end{figure}

\begin{figure}[t!]\centering
	\FloatBarrier
	\includegraphics[trim={0cm 9cm 0cm 0cm},clip,width=\textwidth]{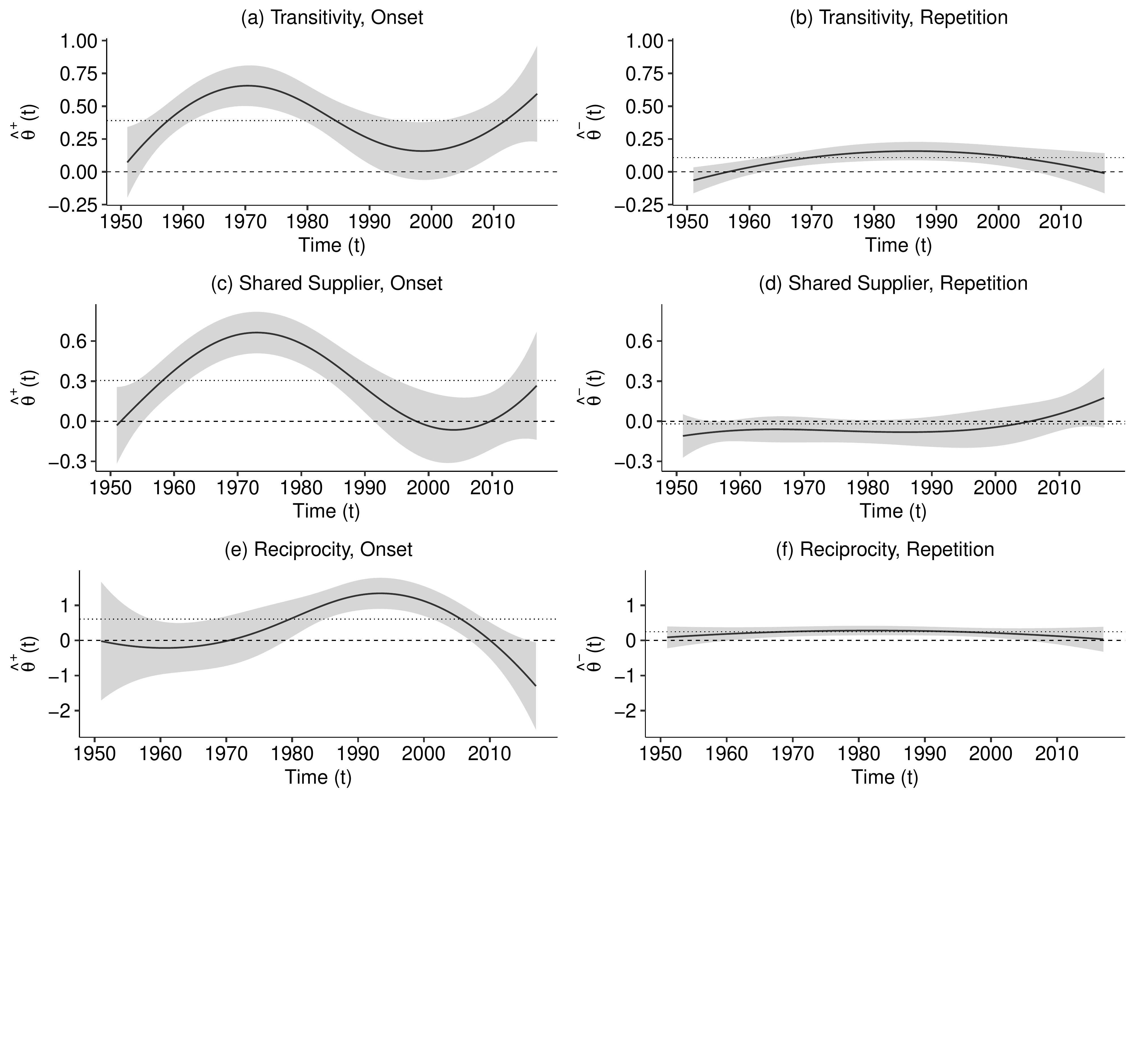}
	\caption{Results of endogenous statistics relating to past dyadic interaction and clustering. The shaded area indicates the $95\%$ confidence bands of the estimates and the dotted horizontal lines represent the time-constant parameters.}
	\label{fig:end2}
\end{figure}

\begin{figure}[t!]\centering
	\FloatBarrier
	\includegraphics[trim={0cm 0cm 0cm 0cm},clip,width=\textwidth]{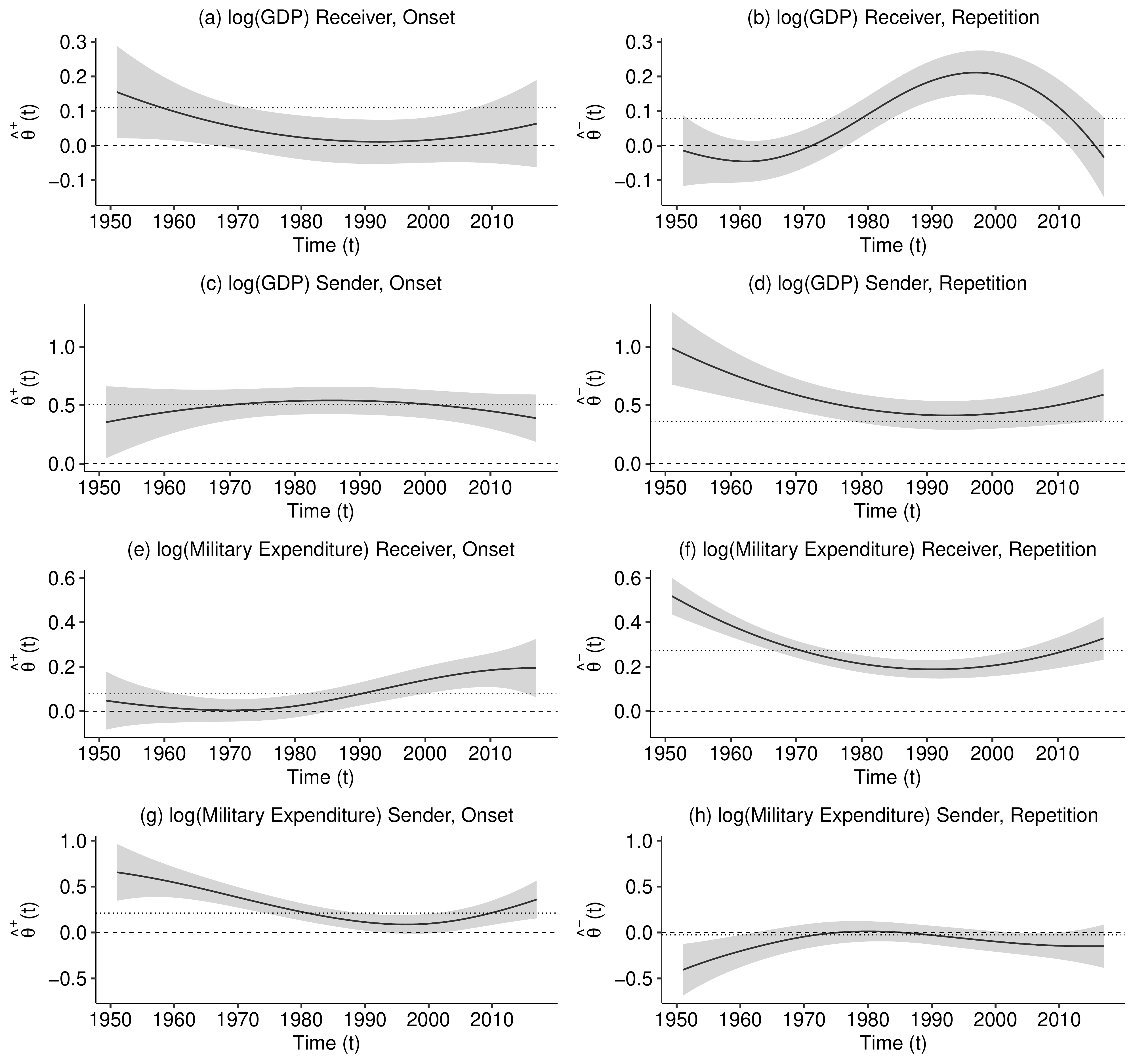}
	\caption{Results of exogenous statistics relating to economic factors. The shaded area indicates the $95\%$ confidence bands of the estimates and the dotted horizontal lines represent the time-constant parameters.}
	\label{fig:exo1}
\end{figure}

\begin{figure}[t!]\centering
	\FloatBarrier
	\includegraphics[trim={0cm 9cm 0cm 0cm},clip,width=\textwidth]{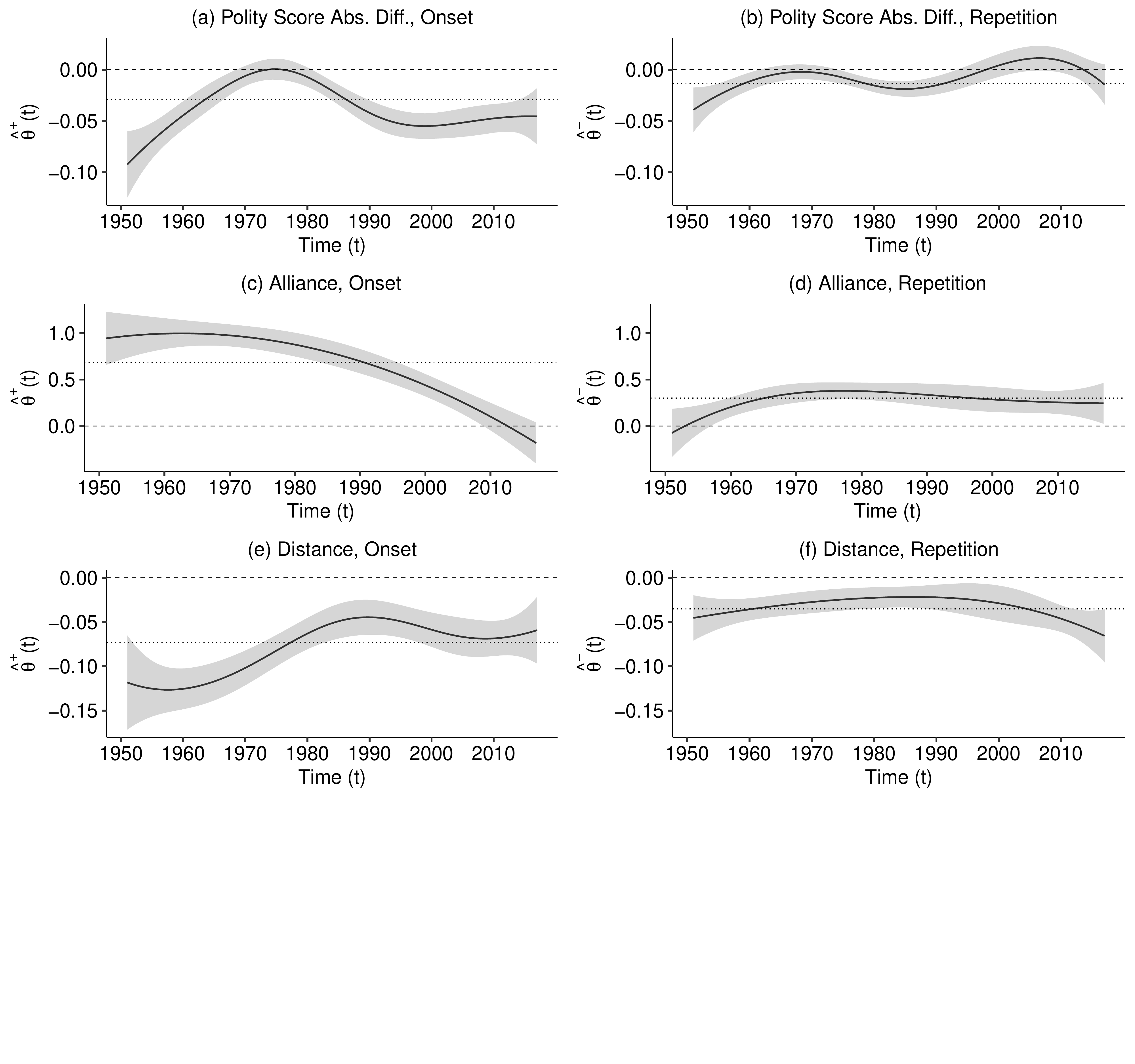}
	\caption{Results of exogenous statistics relating to political, security, and geographical factors. The shaded area indicates the $95\%$ confidence bands of the estimates and the dotted horizontal lines represent the time-constant parameters.}
	\label{fig:exo2}
\end{figure}

In Figures \ref{fig:end1} to \ref{fig:exo2}, the full results of the time-varying estimates are given accompanied by alternative time-constant coefficients as dotted horizontal lines. The latter are obtained by setting $\theta(t) \equiv \theta$. All exponentially transformed estimates at a specific point in time can be interpreted (ceteris paribus) as the multiplicative change of the intensity \eqref{eq:REMseparable} corresponding to the effect of covariates in relative risk models \citep{Kalbfleisch2002}. Therefore, an effect estimated at zero does not change the relative risk of an event to happen, but positive or negative coefficients lead to a higher or lower relative risk of the event to occur, respectively. Additionally, an event's occurrence is equivalent to the increment of one in the counts of aircraft units since one event represents a combat aircraft delivery in our application case. 

From simple inspection, it can be concluded that in all cases, time-varying coefficients are carrying completely different information as compared to time-constant coefficients. This is evidence of the necessity to account for the multiple systemic changes within the international aircraft market during the considered time interval. From a statistical point of view, the time-varying effects can also be underpinned by a lower cAIC value when compared to time-constant effects (see Section \ref{sec:model_assessment} for additional details on the cAIC).   

Moreover, we observe different shapes of the curves of the time-varying coefficients when comparing  \textsl{onset} and  \textsl{repetition} conditions leading to the conclusion that the import of all covariates on these two separate conditions is different. 

Time-varying effects relating to the degree structure are shown in Figure \ref{fig:end1}. Figure \ref{fig:end1} (a) indicates a steady negative influence of the sender`s in-degree in the \textsl{onset} condition from around 1965 onward. It can be concluded that the count of dyadic events is lower if the sender’s in-degree is high, which may be justified by the observation that only a small subset of countries is adequately equipped to produce and export aircraft. This technological possibility, in turn, increases self-sufficient behavior, thus alleviates the need for additional imports. Contrary, in the \textsl{repetition} condition, the in-degree of the receiver exhibits a positive effect for the post-Cold War period from 1990 to 2010, Figure \ref{fig:end1} (b). Otherwise, the effect is insignificant. Concerning the receiver, a negative effect of the in-degree can be observed from 1950 to 1980 in the \textsl{onset} model, Figure \ref{fig:end1} (c). When proceeding to deliver aircraft, the receiver’s in-degree effect is similar to the sender’s in-degree effect, Figure \ref{fig:end1} (d). For the sender’s out-degree, the effect in the \textsl{onset} model is negative until around 1980 and thereupon positive. In the latter case, the effect mirrors a higher tendency of delivering combat aircraft if the sender is already a prolific exporter country. During the entire observational period, we observe that receivers are not senders themselves, thus exhibit low out-degrees, Figure \ref{fig:end1} (g) and (h). This behavior does not depend on the condition of the dyadic intensity. 

\FloatBarrier

The specified triadic structures play a substantial role during the Cold War. Afterwards, the impact disappears but is again strengthened after 2000 under the \textsl{onset} condition, Figure \ref{fig:end2} (a) and (c). In particular, an increasing number of indirect transitive connections between country $i$ and $j$ results in a greater count of aircraft deliveries between 1950 and 1990. Similarly, receiving combat aircraft from the same third country increases the unit sales between the receivers during the Cold War period,  Figure \ref{fig:end2} (c). A possible consequence of this process is the strengthening of a block structure. For a consecutive delivery, the triadic effects are less pronounced, and in the case of shared suppliers,   Figure \ref{fig:end2} (d), constantly insignificant. The count of reciprocal events, on the other hand, raises trade from 1990 to 2005, Figure \ref{fig:end2} (e). This result may be a consequence of an international market opening after the Soviet Union's fall, leading to multiple emergent countries. If the relationship is maintained, reciprocal events are encouraged throughout the period of observation, although to a smaller degree, Figure \ref{fig:end2} (f). 

\FloatBarrier


While the logarithmic GDP of the receiver has a relatively weak positive influence when starting a trade relation, Figure \ref{fig:exo1} (a), its repetition is only affected after the end of the Cold War, Figure \ref{fig:exo1} (b). On the sender-side, the estimates of both models are constantly positive, Figure \ref{fig:exo1} (c) and (d). In contrast to the effect in the onset model, the sender's logarithmic GDP has a higher effect from 1950 to 1980 in the \textsl{repetition} condition. Moreover, the military expenditure of the receiver is one of the main drivers in this model, Figure \ref{fig:exo1} (f). Here, higher military spending of possible sender countries augments the count of receiving combat aircraft deliveries, specifically during the 50s. Conversely, the exogenous covariate only slowly gains attention in the \textsl{onset} condition after the Cold War, Figure \ref{fig:exo1} (e). While the effect of the military expenses of the sender stays overall positive when delivering aircraft for the first time, it inhibits it to be repeated in the next year, Figure \ref{fig:exo1} (g) and (h).

The findings in Figure \ref{fig:exo2} (a) and (b) indicate that similar regimes are overall more likely to start trading combat aircraft. Only at the height of the Cold War from 1970 to 1980, the effect is estimated at approximately 0, Figure \ref{fig:exo2} (a). The strength of the effect is less salient in the \textsl{repetition} condition than in the \textsl{onset} condition of our model, Figure \ref{fig:exo2} (b). Furthermore, the time-varying coefficients discover a steadily decreasing influence of beginning to transact with allies, Figure \ref{fig:exo2} (c).  This finding suggests evidence of the overall deteriorating importance of international alliances in combat aircraft transactions if they did not trade in the previous year.  We don't observe a similar downward trend when repeating an event, Figure \ref{fig:exo2} (d). Lastly, a larger distance between the respective capitals generally hinders events from occurring, Figure \ref{fig:exo2} (e) and (f). Therefore, countries tend to trade with spatially more close than distant partners. This may be caused by the relatively lower transportation cost and is in line with the expectations of the gravity model of trade \citep[see corrigiendum]{johannsen2014,Thurner2019}. 



\FloatBarrier

\subsubsection{Random Effects}
\FloatBarrier

\begin{figure}[!]
	\centering
	\hspace*{0.4cm}	\vspace{-0.9cm}\includegraphics[width=1.2\linewidth]{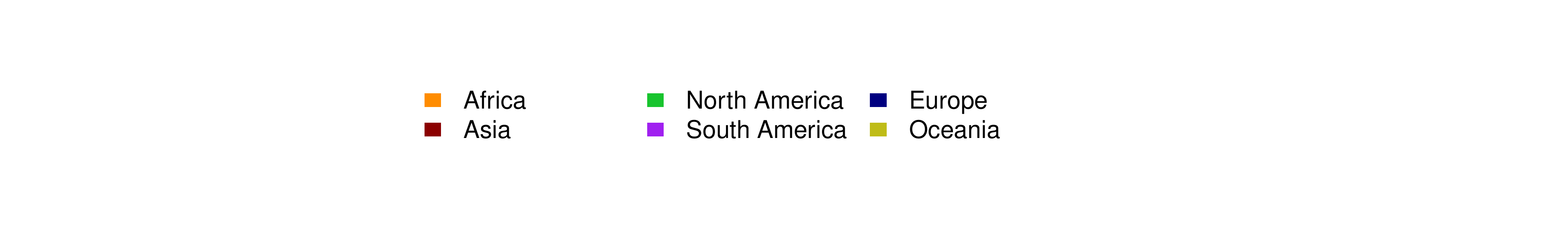} 
	\includegraphics[width=0.55\linewidth]{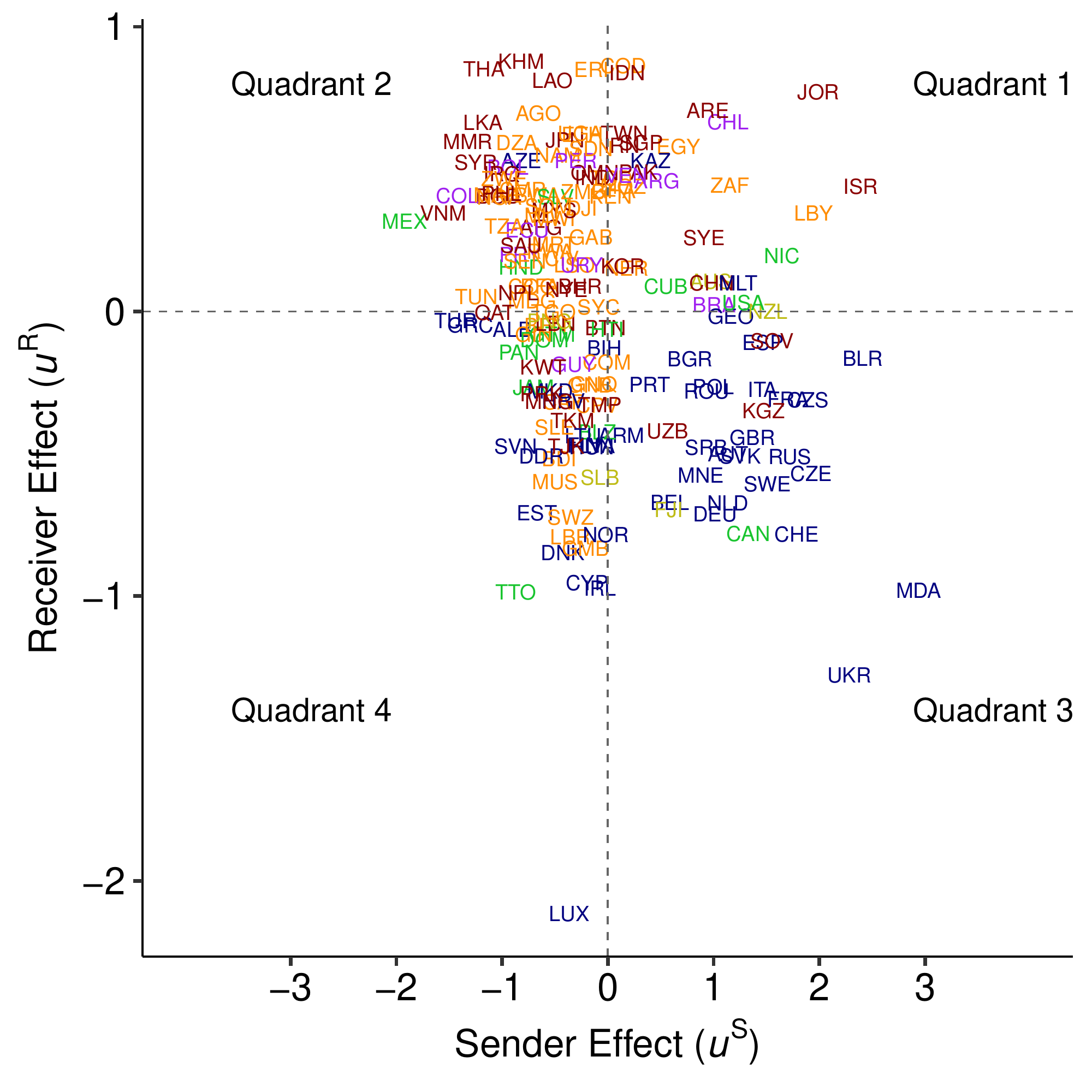}
	\caption{Country-specific random sender and receiver effects. The drawn label represents the respective ISO3 code of the represented country.}
	\label{fig:randomeffects}
\end{figure}

\begin{figure}[t!]
	\centering
	\includegraphics[width=\linewidth]{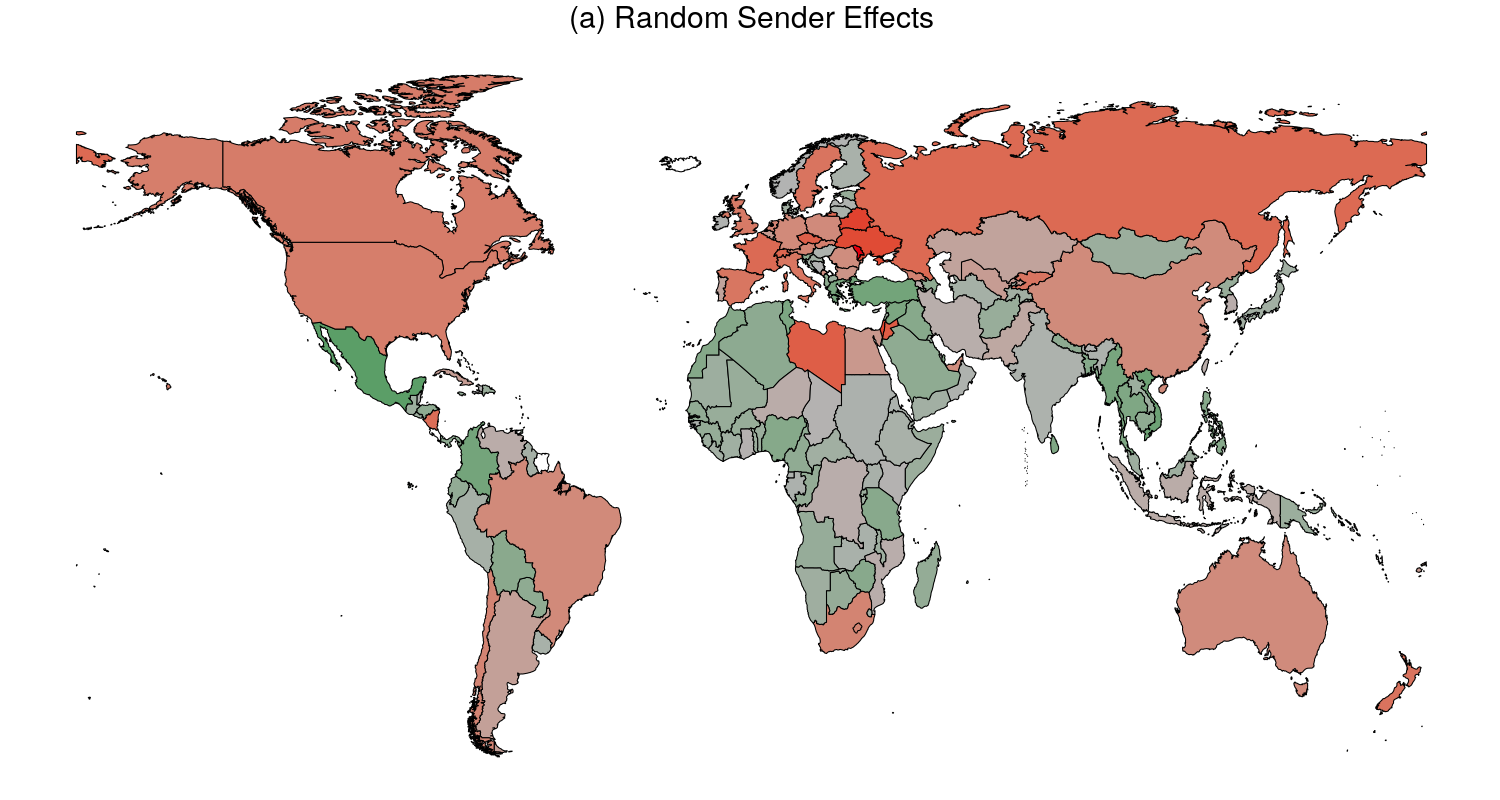}
	\includegraphics[width=\linewidth]{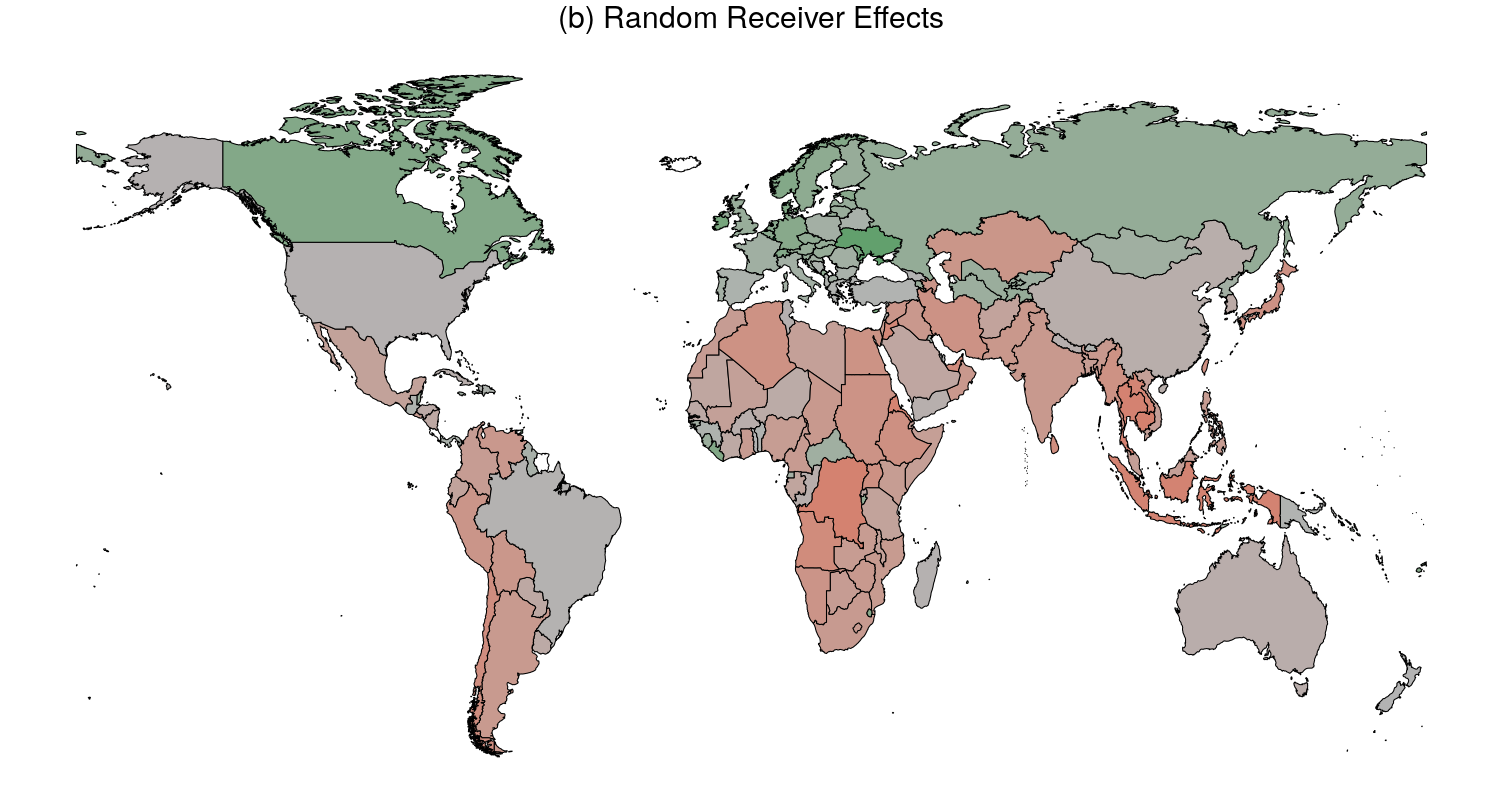}
	\includegraphics[width=\linewidth]{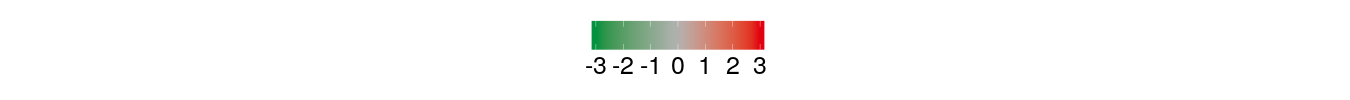}
	\caption{Random country-specific sender (a) and receiver (b) effects. The layout represents the borders as of 2020.}
	\label{fig:maps}
\end{figure}

The random effects permit an extended analysis of the unexplained heterogeneity in the model. More precisely, the random effects express country-specific deviations from an overall behavioral trend captured by the time-varying effects. Additionally, they correct the countries' repeated measurements as simultaneous senders and receivers of events in each year. The model introduced in Section \ref{sec:Process-based Model} comprises two country-specific random effects for all countries as a sender and receiver of combat aircraft deliveries. The results are given in Figure \ref{fig:randomeffects} and visualized on a world map in Figure \ref{fig:maps}.

In the first quadrant of Figure \ref{fig:randomeffects} countries with a positive random sender and receiver effect are shown. This composition of random effects suggests that the respective countries are senders and receivers of more combat aircraft events than marginally expected. Countries in the Middle East, e.g., Israel (ISR), Libya (LBY), and Jordanian (JOR), are allocated to this group. 

Negative sender but positive receiver effects are identified for countries in South-East Asia (Thailand (THA), Cambodia (KHM), Laos (LAO), Myanmar (MYR), and Sri Lanka (LKA)). Compared to the average behavior, these countries are somewhat reluctant as senders and confident as receivers of combat aircraft deliveries. The latent sender effect of Mexico (MEX) is the most negative coefficient estimated. This suggests Mexico's reliance on the import of combat aircraft, although its high economic status would imply additional participation in the event network as a sender.

The third quadrant contains all countries, which were less active than expected as a sender and receiver of events. This strand of countries is either economically strong, yet exhibiting a passive trading behavior, e.g., Luxembourg (LUX), or relatively poor and missing preconditions to send or receive weapons, e.g., Trinidad and Tobago (TTO).

Lastly, a negative random coefficient regarding receiving arms is mostly associated with European countries. The corresponding sender effect is positive. Hence, these countries are situated in the fourth quadrant of Figure \ref{fig:randomeffects}.  The East European countries Moldova (MDA), Ukraine (UKR), and Belarus (BLR) have the highest positive sender effect paired with relatively low receiver effects. 

In terms of continent-wide tendencies, we locate Africa in the first three quadrants. South America is principally assigned to the first and second quadrant. Asia, Oceania, and North America are more dispersed and exhibit less homogeneous country behavior.  
\FloatBarrier
\subsection{Model Comparison and Assessment}
\label{sec:model_assessment}
We compare the estimated model to alternative specifications, which are chosen to reflect all subsequent extensions of Section \ref{sec:Extensions} and are indicated in Table \ref{tbl:specs}. Model 1 includes all effects linearly without the separable extension. This is we assume that $\theta(t) \equiv \theta$ and omit the separation of the statistics $	s_{ij}(y_{t-1}, x_{t-1})$ into $s_{ij}^{+}(y_{t-1}, x_{t-1})$ and $s_{ij}^{-}(y_{t-1}, x_{t-1})$. This separability is added in Model 2 according to Section \ref{sec:Separability Assumption}. Model 3 includes time-varying coefficients as introduced in Section \ref{sec:Time-Varying Effects}. Lastly, Model 4 is the model whose findings were presented in Section \ref{sec:res}. Hence, also random effects are taken into account, that are explained in Section \ref{sec:Accounting for Nodal Heterogeneity}. 

	\begin{table}[!]
			\caption{Specifications of the compared models and resulting corrected AIC (cAIC)  values.}
		\label{tbl:specs}
	\center
	\begin{tabular}{c c c c c c} 
		& Separability & Time-Varying Effects & Random Effects &   cAIC    \\ \hline
		Model 1 & \xmark    &       \xmark      &     \xmark     & 84622.47  \\
		Model 2 &  $\checkmark$  &        \xmark        &    \xmark     & 65614.86  \\
		Model 3 &  $\checkmark$  &        $\checkmark$        &   \xmark   & 63174.49 \\
		Model 4 &  $\checkmark$  &      $\checkmark$      &   $\checkmark$   & 59717.77
	\end{tabular} 

\end{table}

One way to compare these models is by means of information criteria, i.e. the Akaike Information Criterion (AIC,\citealp{Akaike1974}). As already discussed in the context of linear mixed models \citep{Greven2019} and generalized mixed models \citep{Saefken2014}, the usage of the conditional or marginal AIC does not appropriately incorporate the uncertainty of estimating the covariance parameters of the random effects (in our application $\tau_S^2$ and $\tau_R^2$). Therefore, we utilize a corrected conditional AIC proposed by \citet{wood2016}. The resulting cAIC values are given in Table \ref{tbl:specs} and indicate a superior model fit when all extensions introduced in Section \ref{sec:Extensions} are included. 

\begin{figure}[t!]
	\centering
	\includegraphics[width=0.6\linewidth]{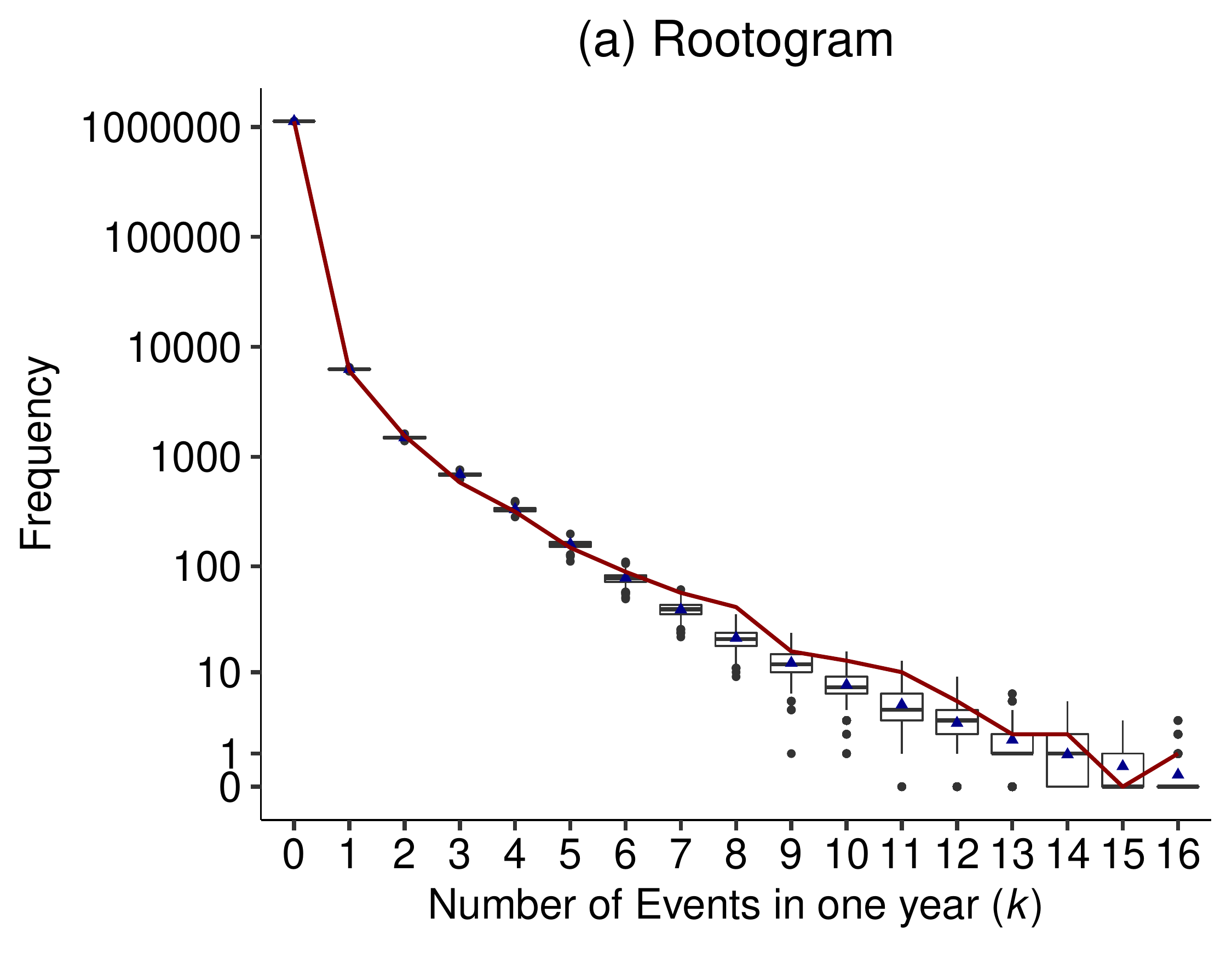}
	\includegraphics[width=0.45\linewidth]{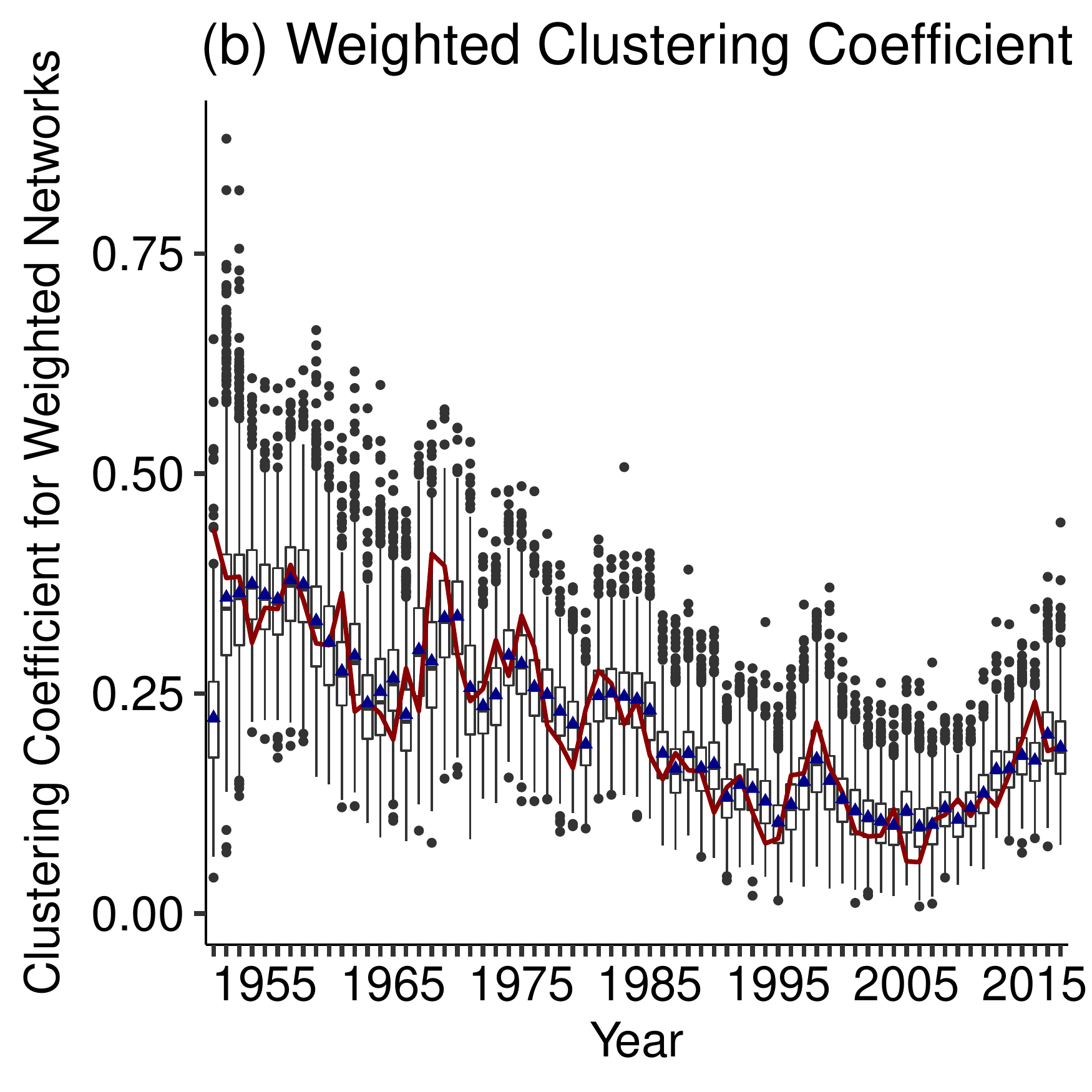}
	\includegraphics[width=0.45\linewidth]{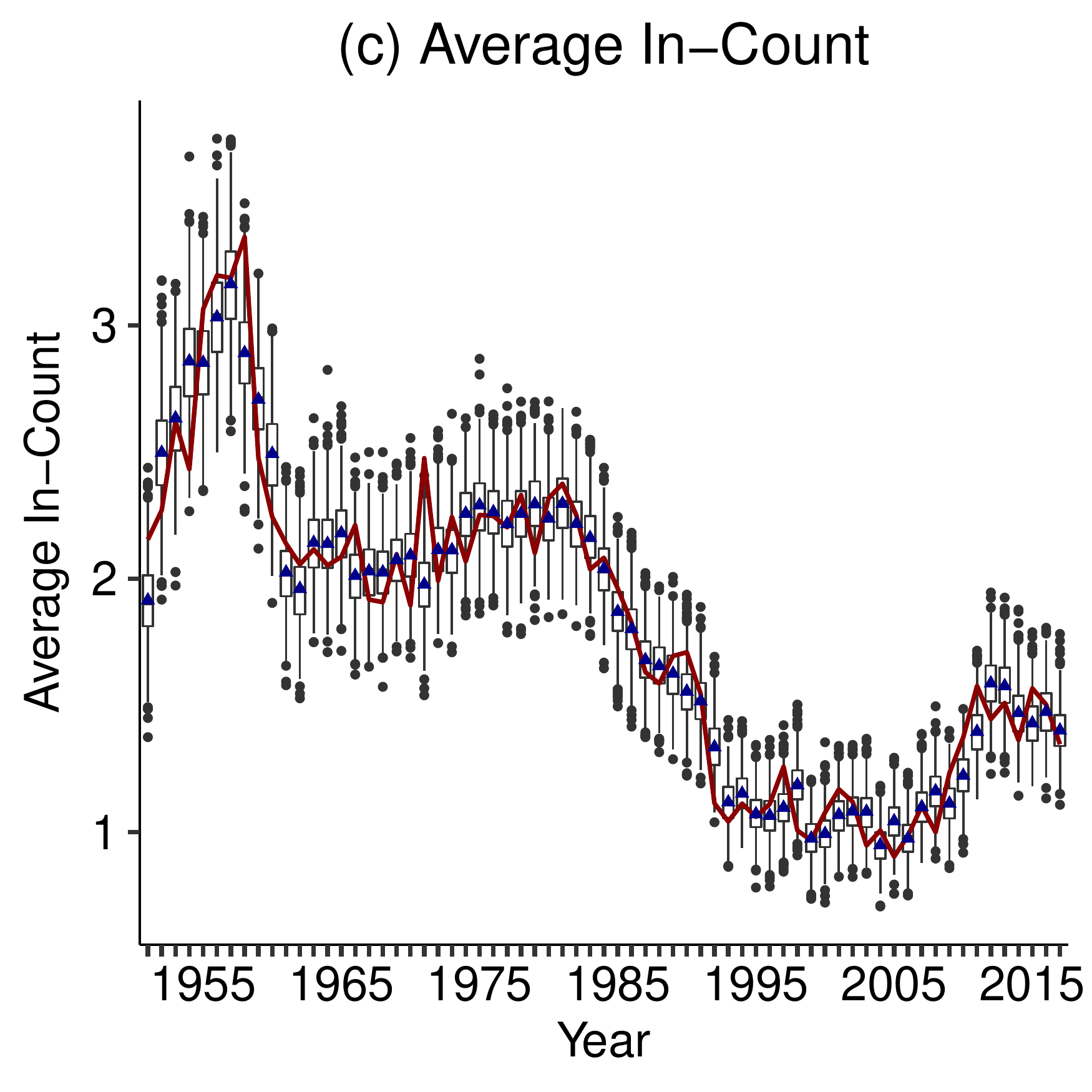}
	\caption{Comparison of the observed and simulated frequencies of the dyadic event counts (a), weighted clustering coefficients over time (b), and average in-count over time (c). The red lines indicate the observed values of each statistic, whereas the boxplots are the result of drawing 1000 networks according to \eqref{eq:assumption_random} and the blue triangles the average values. }
	\label{fig:modelassesmentrootogram}
\end{figure}

We assess the selected Model 4 with a graphical tool proposed by \citet{Hunter2008b} for general network models. The procedure's basic idea is to evaluate whether networks randomly generated according to the estimated network model at hand conserve pre-specified characteristics of the observed network reasonably well. In our particular case, we simulate yearly increments of our network counting process from \eqref{eq:assumption_random} and consider the result as a count-valued network. However, most network statistics commonly used for this assessment are solely defined for binary networks. Therefore, we propose a suite of novel statistics for our application case. To detect whether our model adequately replicates possible over- or underdispersion in the count data, we rely on the statistics from rootograms, i.e., the empirical and simulated frequencies of the counts in the networks. For general regression tasks involving count data, rootograms were proposed by \citet{Kleiber2016a} and date back to \citet{Tukey1977}. Usually, one compares the square-root-transformed observed and expected frequencies of the target variable. However, in our application, we substitute the square-root transformation with a log transformation due to the high percentage of zeros and use the simulated rather than expected frequencies to fit into the framework of \citet{Hunter2008b}. Secondly, we investigate to what extent the performance of our model is stable over the time frame we analyze.  To do so, we compute the clustering coefficient for weighted networks as proposed by \citet{Opsahl2009}\footnote{We opt for the variant of the statistic that aggregates triplets of event counts within a year via the arithmetic mean.} for the yearly networks $\mathbf{y}_t$. Besides, we examine the average in-count per year, which is directly related to the average count per year. In the Supplementary Material, we show how the distribution of the observed count distributions of in- and outgoing events given in Figure \ref{fig:outdeg2} are reproduced in the simulated networks and provide the mathematical formulations of all statistics. 

Figure \ref{fig:modelassesmentrootogram} shows the variability of all specified statistics computed for all 1000 simulated networks through boxplots and displays the average value by a blue triangle. Red lines indicate the observed measurements. We can infer from Figure \ref{fig:modelassesmentrootogram} (a) that the estimated model captures even high event counts between countries averaged over the entire period. At the same time,  our proposed model is capable of representing the yearly clustering as well as the average in-count, see Figure \ref{fig:modelassesmentrootogram} (b) and (c). Therefore, we gather that the performance of the proposed model is consistently good throughout the observational period. 

%



\section{Conclusion}
\label{sec:conclusion}

We introduced a novel model for the analysis of relational event data. Originating in a counting process operating in continuous time that we only observe at specific time points, we derived a tie-level intensity, whose parameters can be estimated according to the maximum likelihood principle. Extensions to separable models, which govern the \textsl{onset} and \textsl{repetition} of events by two functions, and the incorporation of time-varying and random coefficients are given. Eventually, we applied the procedure to the international combat aircraft network from 1950 to 2017. By doing that, we use the additional information provided by the counts of yearly aircraft deliveries to estimate a time-continuous intensity, contrary to existing work on binarized networks. Moreover, the separability detects fundamentally different processes governing the \textsl{onset} and \textsl{repetition} of event relationships, while the time-varying effects uncover a systemic change during the Cold War period. Furthermore, we identified triangular network statistics and the sender's economic nodal covariates as the principal drivers of the \textsl{onset} condition of the proposed intensity. Here, a decaying effect of bilateral military alliances became apparent. For the \textsl{repetition} condition, this effect remained consistently positive, and the receiver's high military expenditure was shown to be the driving force. Finally, the random effects enable a visual comparison of the unexplained heterogeneity between the modeled countries (Figure \ref{fig:maps}) and correct the estimates for repeated measurements as well as possible overdispersion. 

	\section*{Acknowledgment}
	We thank the anonymous reviewers for their careful reading and constructive comments. The project was supported by the European Cooperation in Science and Technology [COST Action CA15109 (COSTNET)]. We also gratefully acknowledge funding provided by the German Research Foundation (DFG) for the project  KA 1188/10-1 and TH 697/9-1: \textit{International Trade of Arms: A Network Approach}. Furthermore, the work has been partially supported by the German Federal Ministry of Education and Research (BMBF) under Grant No. 01IS18036A. The authors of this work take full responsibilities for its content.

	\section*{Conflicts of Interest}
	
All authors have nothing to disclose.

\bibliographystyle{apa}
\bibliography{library}

\label{lastpage}
\end{document}


\maketitle
\appendix

\tableofcontents
\newpage

\section{Correlation of Endogenous Statistics between Subsequent Years}

\begin{figure}[!]
	\FloatBarrier
	\centering
	\includegraphics[width=0.6\linewidth]{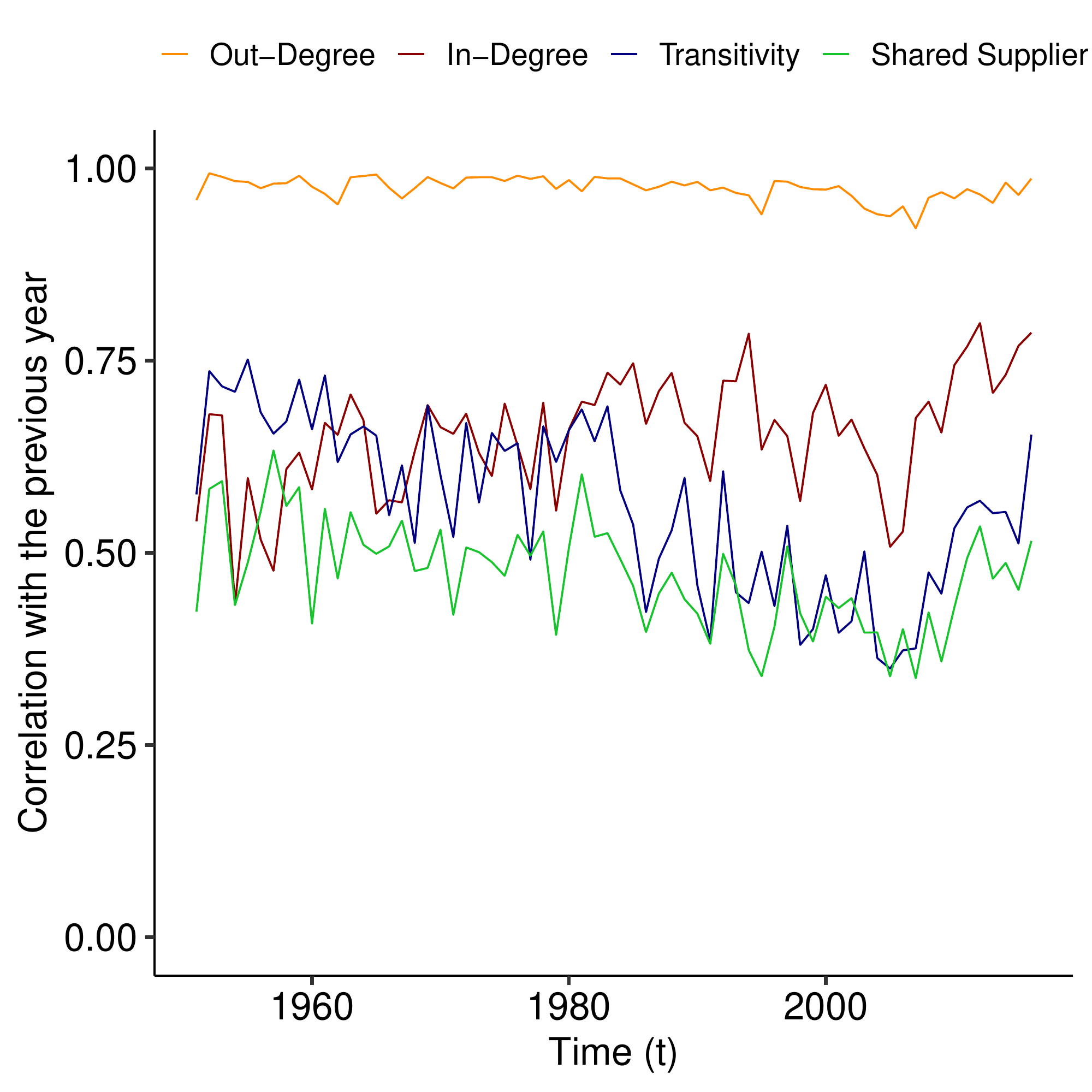}
	\caption{Yearly auto-correlations of endogenous covariates. }
	\label{fig:corr}
\end{figure}

To further legitimize the usage of lagged endogenous covariates, we investigate the yearly auto-correlations of the corresponding statistics. Therefore, we construct time series on the monadic level for the in- and out-degrees of each country and at the dyadic level for triangular statistics, i.e., regarding a tuple of countries. In Figure \ref{fig:corr}, we then descriptively analyze the yearly correlation between all statistics, where measurements are available at both time points. The results again highlight the reliability of using the endogenous statistics of the past year as a proxy for the current year, as we observe exceptionally high correlations. 

\newpage 

\section{Data Sources}
\FloatBarrier

\begin{table}[th!]
	\center
	\caption{Data sources of the exogenous covariates. Versions are indicated where available. }
	\label{tbl:data_source}
	\begin{tabular}{p{7cm} p{1cm}  p{1cm}  p{4.5cm}}
		Covariate & From & To & Data Source \\ 
		\hline 
		\multirow{2}{*}{GDP, Base-Year 2005}    & 1950 &2011 & \citet{GLEDITSCH2002}, v4.1\\ 
		& 2012 & 2017   &  \citet{GDP2017} \\ \hline
		\multirow{2}{*}{Military Expenditure, Base-Year 2017}    & 1950 &2000  &  \citet{singer1972}, v5.0 \\ 
		& 2000 &2017  &  \citet{SIPRI}\\ \hline
		Polity Score   & 1950 &2017  & \citet{marshall2017} \\ \hline
		Alliance & 1950 &2017  &  \citet{Leeds2019}, v4.01\\ \hline
		Distance of Capitals  & 1950 &2017  & \citet{gleditsch2013} \\ 
		
	\end{tabular} 
	
\end{table}

\FloatBarrier

\section{Further Descriptive Analysis}

\FloatBarrier

\begin{figure}[!]
	\FloatBarrier
	\centering
	\includegraphics[width=0.48\linewidth]{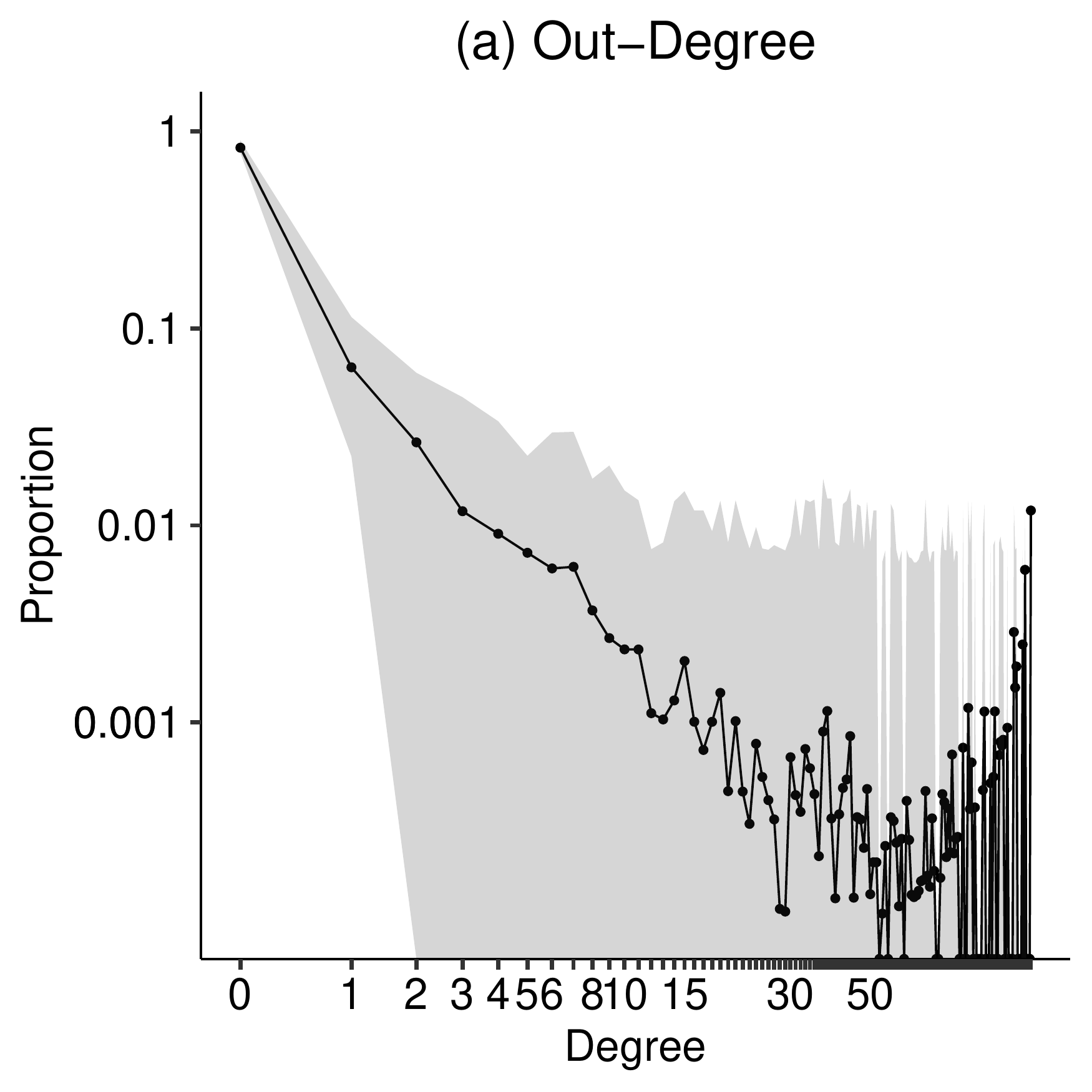}
	\includegraphics[width=0.48\linewidth]{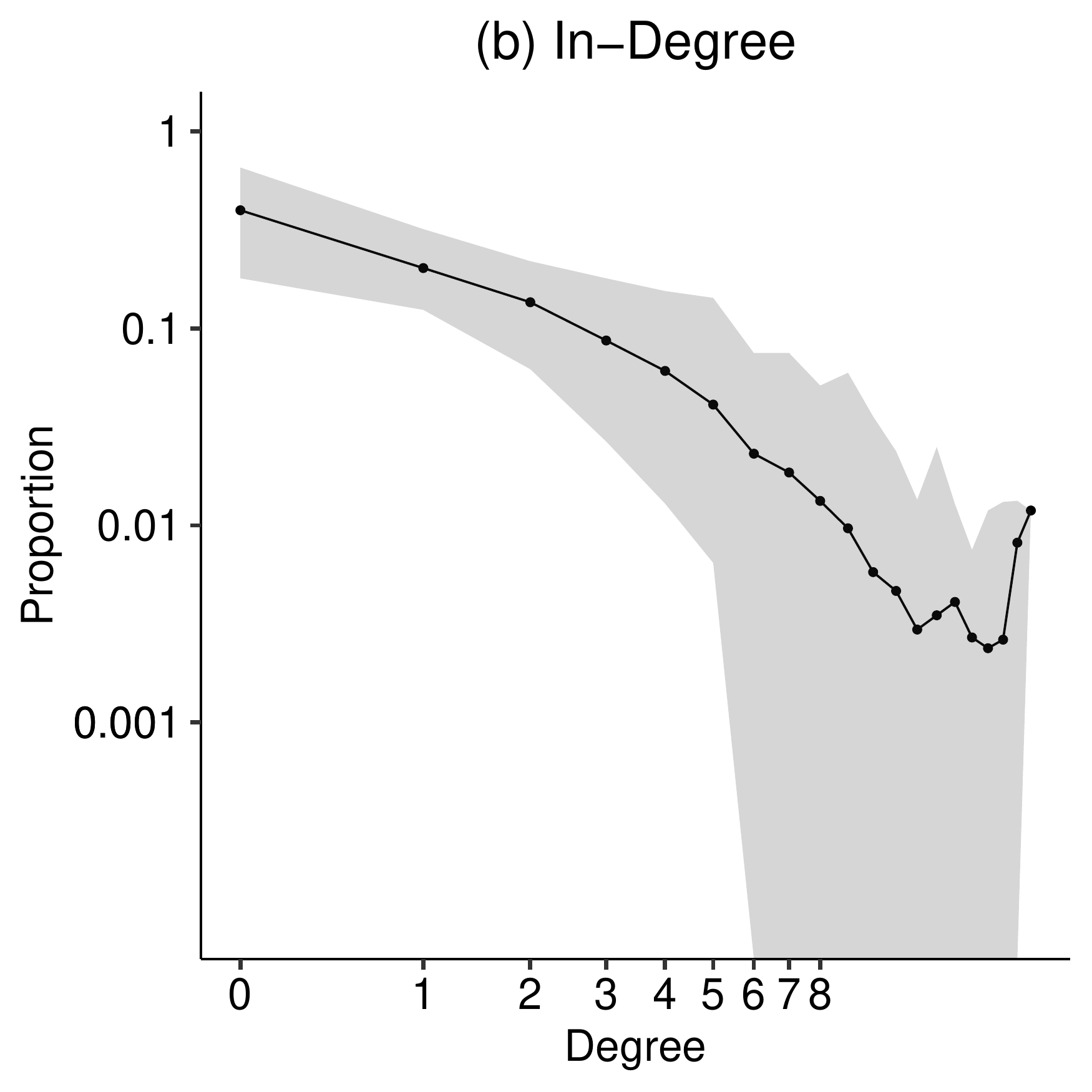}
	\caption{Average Degree Distributions of the Out- and In-Degree for all included countries. The shaded area represents the minimum and maximum observed value. All graphs are represented on a logarithmic scale.}
	\label{fig:outdeg}
\end{figure}

The distribution of the in- and out-degrees can be used to analyze the topology of general networks \citep{barabasi1999,Snijders2003,Newman2002}.  Similar to the findings in Figure 2 of the main article,   \ref{fig:outdeg} (a) underpins the strong centralization of the out-degree distribution. Again mirroring the results of the main article, the in-degree distribution is not as skewed, Figure \ref{fig:outdeg} (b). There are few high degree countries, but the mode is still at zero.

\begin{figure}[ht!]
	\FloatBarrier
	\centering
	\includegraphics[width=\linewidth]{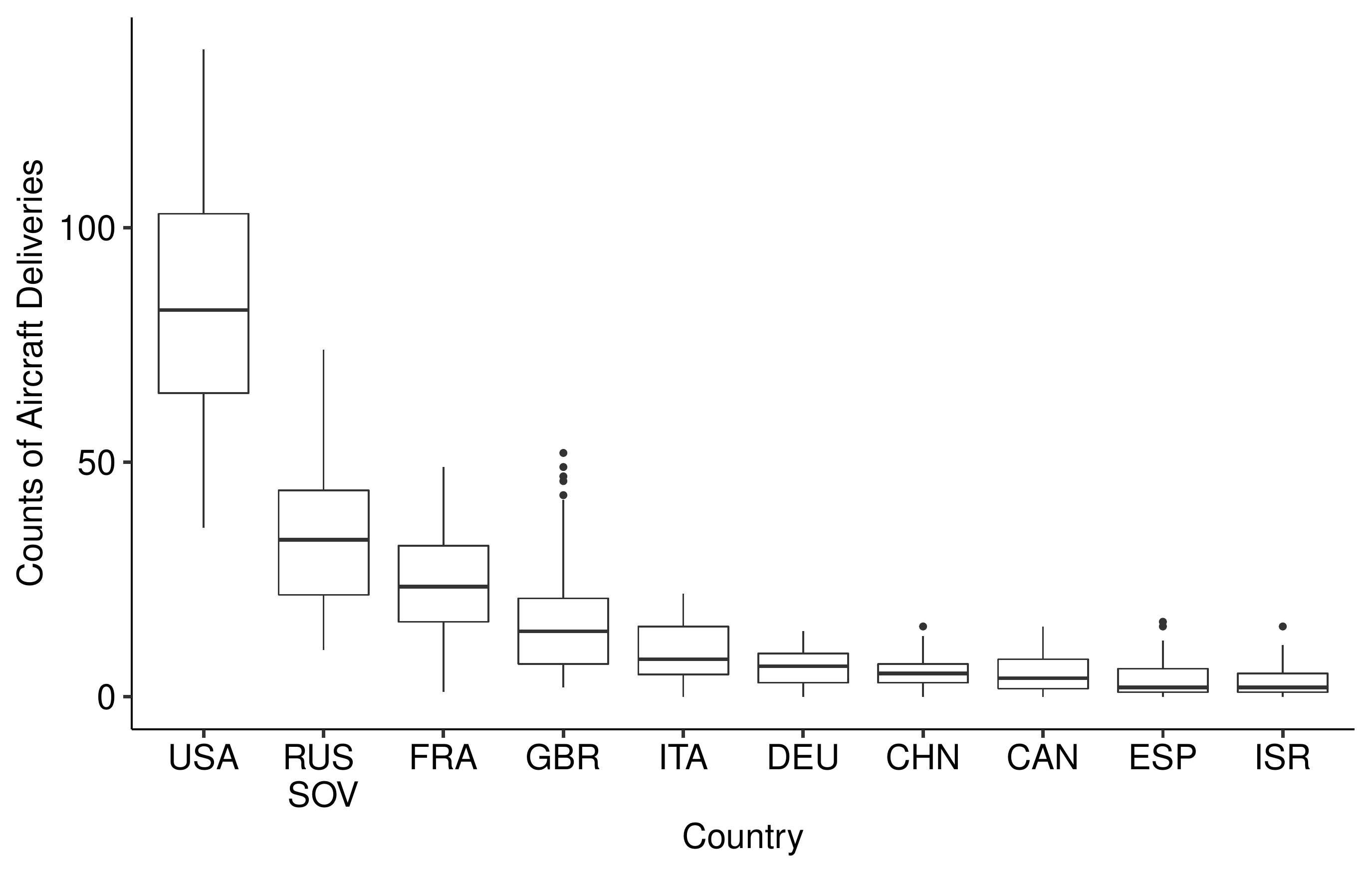}
	\caption{Boxplot of the observed counts over the years of the top 10 sender countries. The labels are the ISO3 codes of the respective countries.}
	\label{fig:boxplot_sender}
\end{figure}

\begin{figure}[ht!]
	\FloatBarrier
	\centering
	\includegraphics[width=\linewidth]{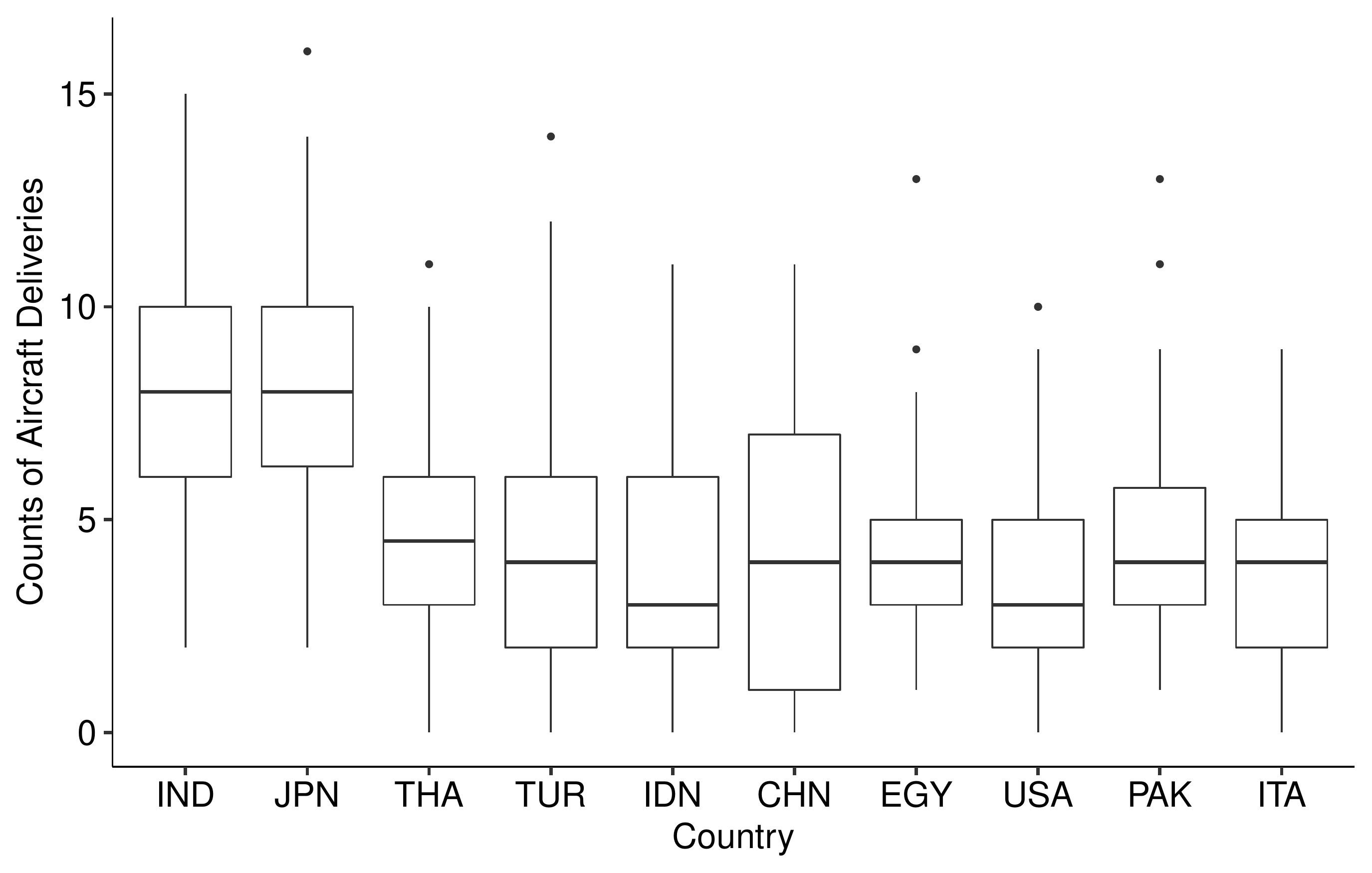}
	\caption{Boxplot of the observed counts over the years of the top 10 sender countries. The labels are the ISO3 codes of the respective countries.}
	\label{fig:boxplot_receiver}
\end{figure}

Alternatively, we can focus the descriptive analysis on the top 10 sender and receiver in the network. The yearly counts of the respective countries are represented as boxplots in Figure \ref{fig:boxplot_sender} and \ref{fig:boxplot_receiver}. The exposed situation of USA is clearly visible, especially in Figure \ref{fig:boxplot_sender}. This role was already thoroughly analyzed in  \citet{Lorell2003}. India predominantly buys combat aircraft from Great Britain, which reflects the dyadic colonial history. Japan, on the other hand, obtains 95$\%$ of the delivered aircraft from USA, being the second highest receiving country.

\section{Robustness Checks}

\subsection{Weighted Fit}

Each event can be comprehended as having a weight given by its TIV. As most possible events in out application were not realized, the respective TIVs are set to zero. Therefore, the weight of the tuple between country $i$ and $j$ at time point $t$ is given by $w_{ij}(t)\propto \log \big(\text{TIV}_{ij}(t) + 1\big) +1$, where  $\text{TIV}_{ij}(t)$ denotes the aggregated TIVs of the same country tuple in the year $t$. The proportionality stems from the fact, that the weights are subsequently standardized so that their sum equals 1.

\begin{figure}[ht!]\centering
	\FloatBarrier
	\includegraphics[trim={0cm 0cm 0cm 0cm},clip,width=\textwidth]{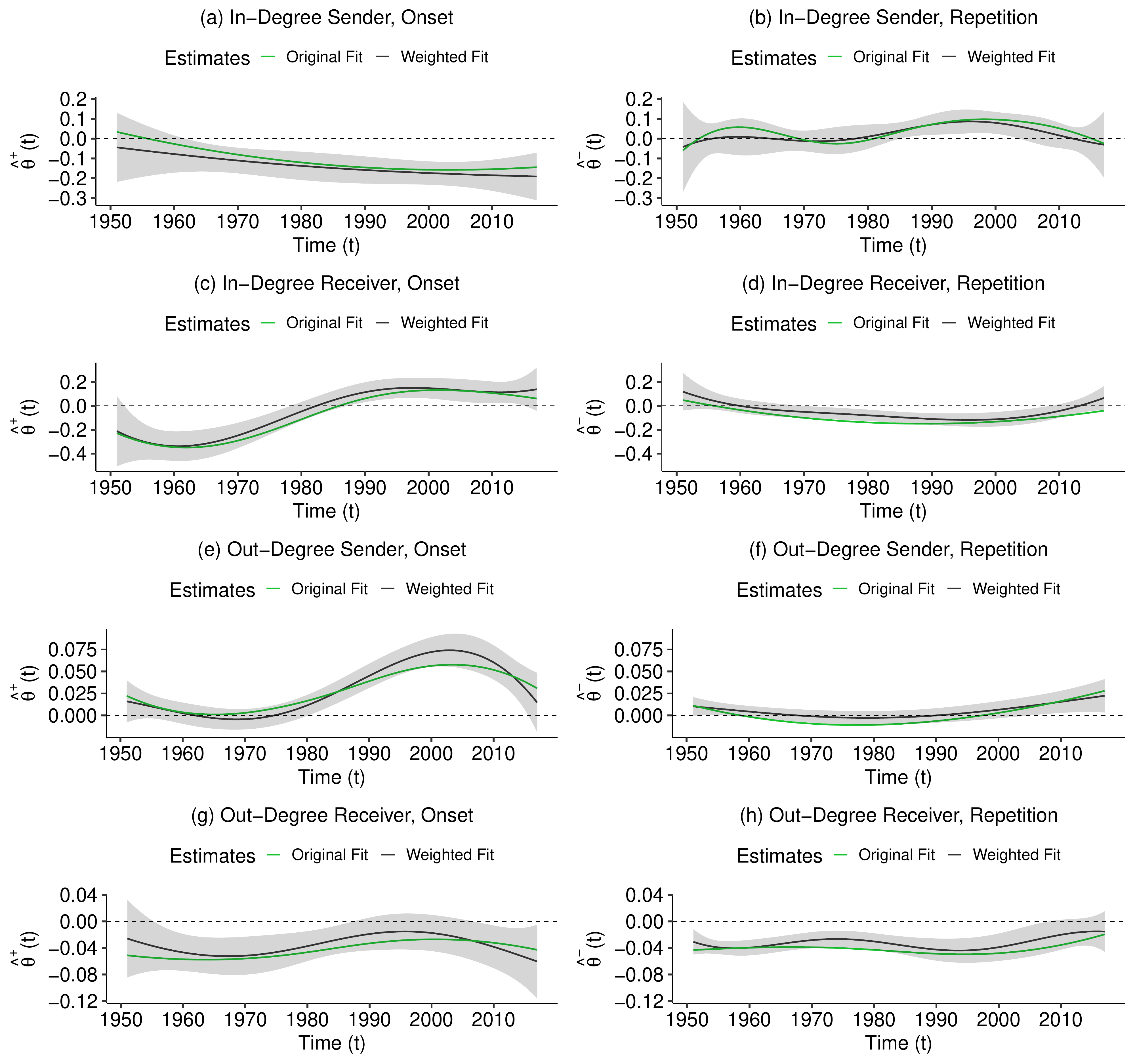}
	\caption{Robustness checks of the estimated parameters comparing the original fit to the model that weighted the observations according to the respective TIV.The green line represents the original fit, while the shaded area indicates the $95\%$ quantile confidence bands of the weighted estimation.}
	\label{fig:weight_end1}
\end{figure}

\begin{figure}[ht!]\centering
	\FloatBarrier
	\includegraphics[trim={0cm 0cm 0cm 0cm},clip,width=\textwidth]{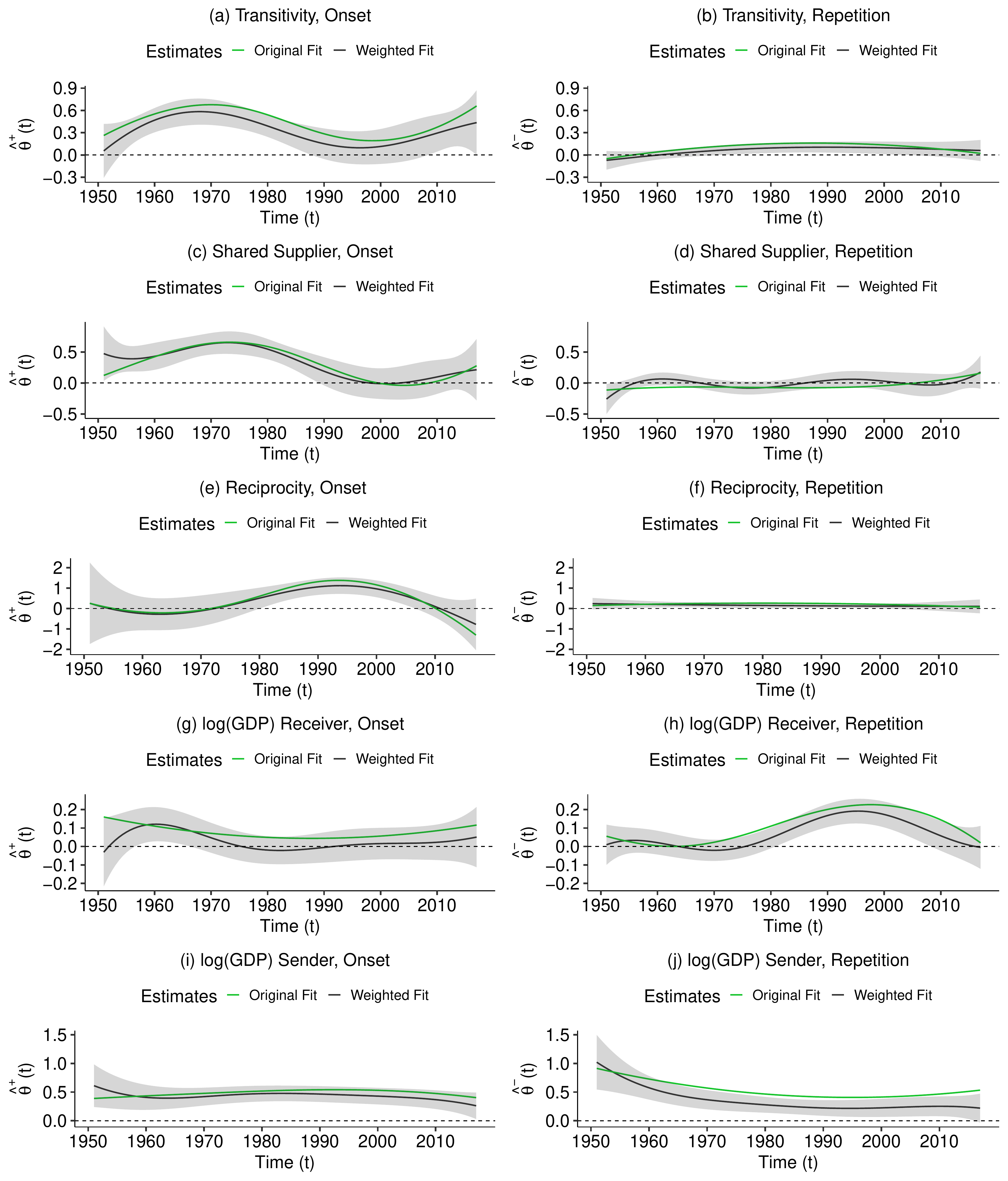}
	\caption{Robustness checks of the estimated parameters comparing the original fit to the model that weighted the observations according to the respective TIV.The green line represents the original fit, while the shaded area indicates the $95\%$ quantile confidence bands of the weighted estimation.}
	\label{fig:weight_end2}
\end{figure}

\begin{figure}[ht!]\centering
	\FloatBarrier
	\includegraphics[trim={0cm 0cm 0cm 0cm},clip,width=\textwidth]{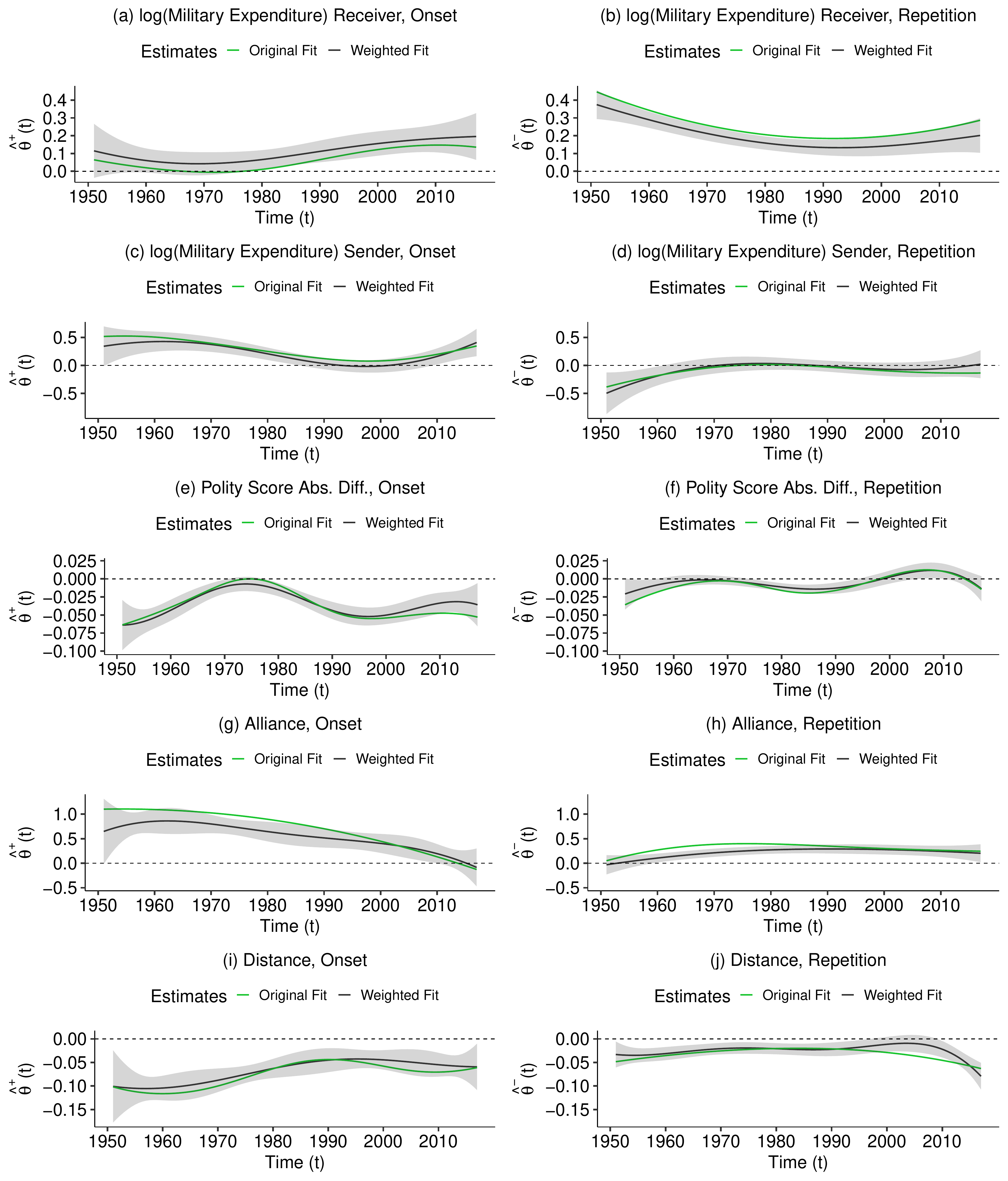}
	\caption{Robustness checks of the estimated parameters comparing the original fit to the model that weighted the observations according to the respective TIV.The green line represents the original fit, while the shaded area indicates the $95\%$ quantile confidence bands of the weighted estimation.}
	\label{fig:weight_exo1}
\end{figure}

Figures \ref{fig:weight_end1}, \ref{fig:weight_end2}, and \ref{fig:weight_exo1} contrast the estimates resulting from the original and weighted fit. The substantial conclusions drawn in Section 4 of the main article are paralleled by the weighted estimates. 

\FloatBarrier

\newpage
~ 
\newpage
~ 
\newpage

\subsection{Alternative Time-Spans defining Separability}
\label{sec:ann_sep}
\FloatBarrier

\begin{figure}[ht!]
	\centering
	\includegraphics[width=0.48\linewidth]{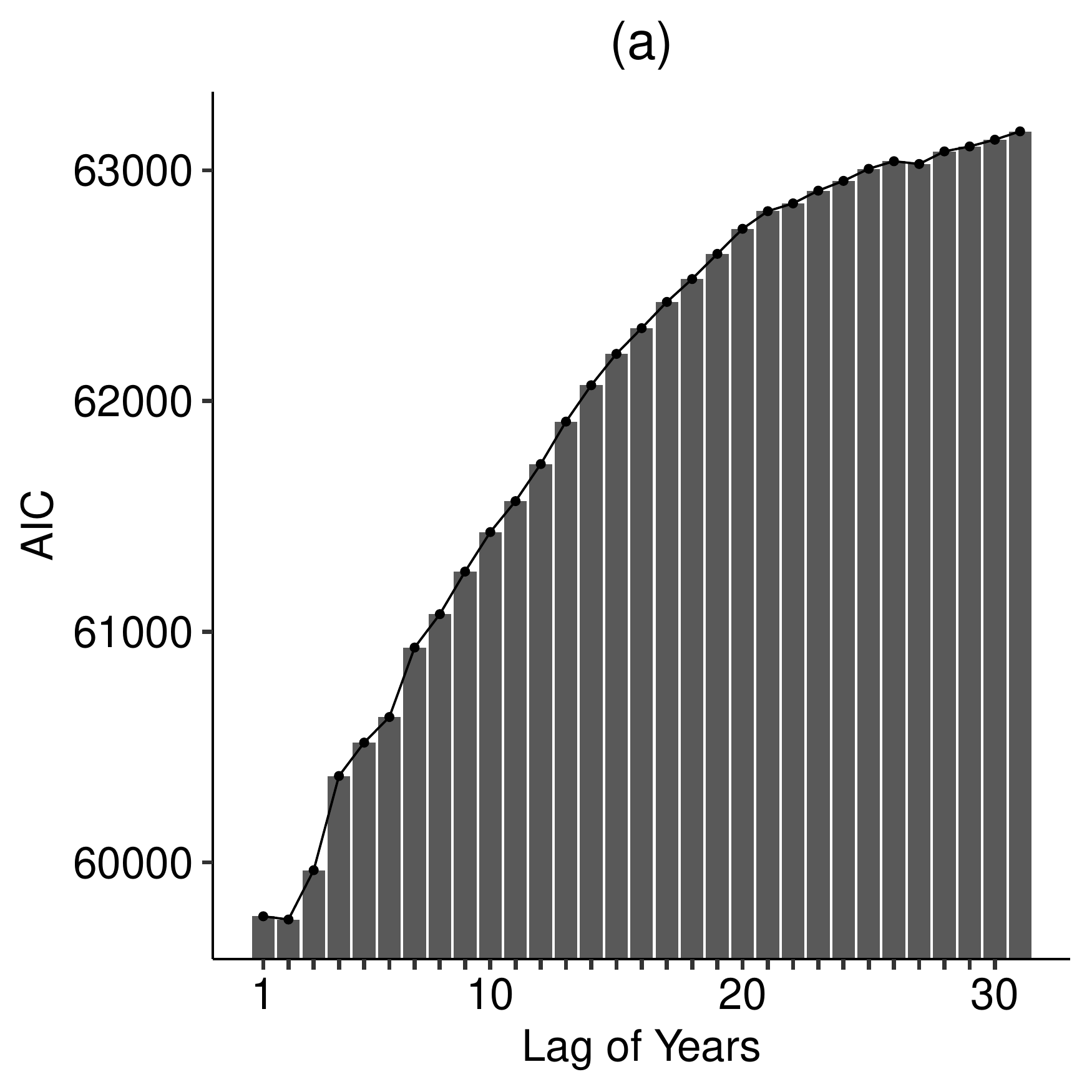}
	\includegraphics[width=0.48\linewidth]{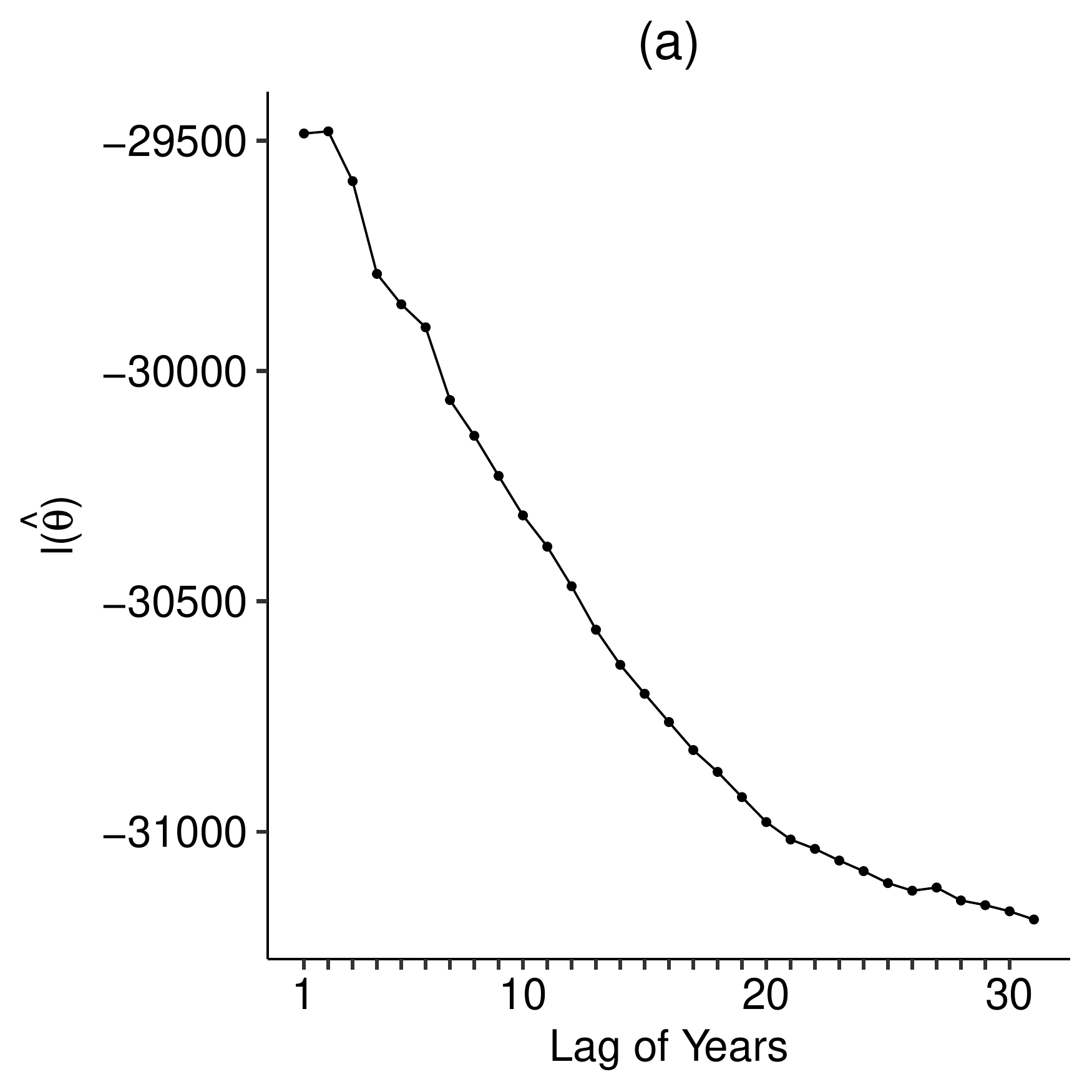}
	\caption{(a): Resulting AIC value by varying the length of the interval defining the separability. (b): The value of the log likelihood evaluated at the final estimates of the respective models.}
	\label{fig:separability}
\end{figure}

The separability assumption can be adapted by changing the time frame, dictating which intensity governs which event. In the application case we fixed this interval to be one year. In order to legitimize this decision, we estimated the exact same model with a varying interval length defining from when an event tuple is, e.g., driven by the \textsl{onset} intensity. For instance, a lag of 10 years would translate to being driving by the \textsl{onset} intensity if two countries did not trade with each other in the last 10 years. Figure \ref{fig:separability} plots the AIC scores and values of the log likelihood evaluated at the final estimates of the respective models over the lag. Apparently, there are only slight differences between using a log of one or two years, yet longer lags lead to a steadily deteriorating performance of the model.

\FloatBarrier

\newpage

\FloatBarrier

\subsection{Thresholds for TIV of Events}

In the application of Section 3 all events were regarded unconditional of their extent. Alternatively, one may only include events above a certain threshold in terms of TIVs of the events. As a robustness check of the findings in the article, we, therefore, repeat the parameter estimation in three different scenarios, which are defined as follows: 

\begin{enumerate}[leftmargin=1cm]
	\item Include events, if their TIV is above the 0.05 quantile of all TIVs ($> z_{0.05}$)
	\item Include events, if their TIV is above the 0.1 quantile of all TIVs ($> z_{0.1}$)
	\item Include  events, if their TIV is above the 0.15 quantile of all TIVs ($> z_{0.15}$)
	\item Include all events (Full Data)
\end{enumerate}

\begin{figure}[ht!]\centering
	\FloatBarrier
	\includegraphics[trim={0cm 0cm 0cm 0cm},clip,width=\textwidth]{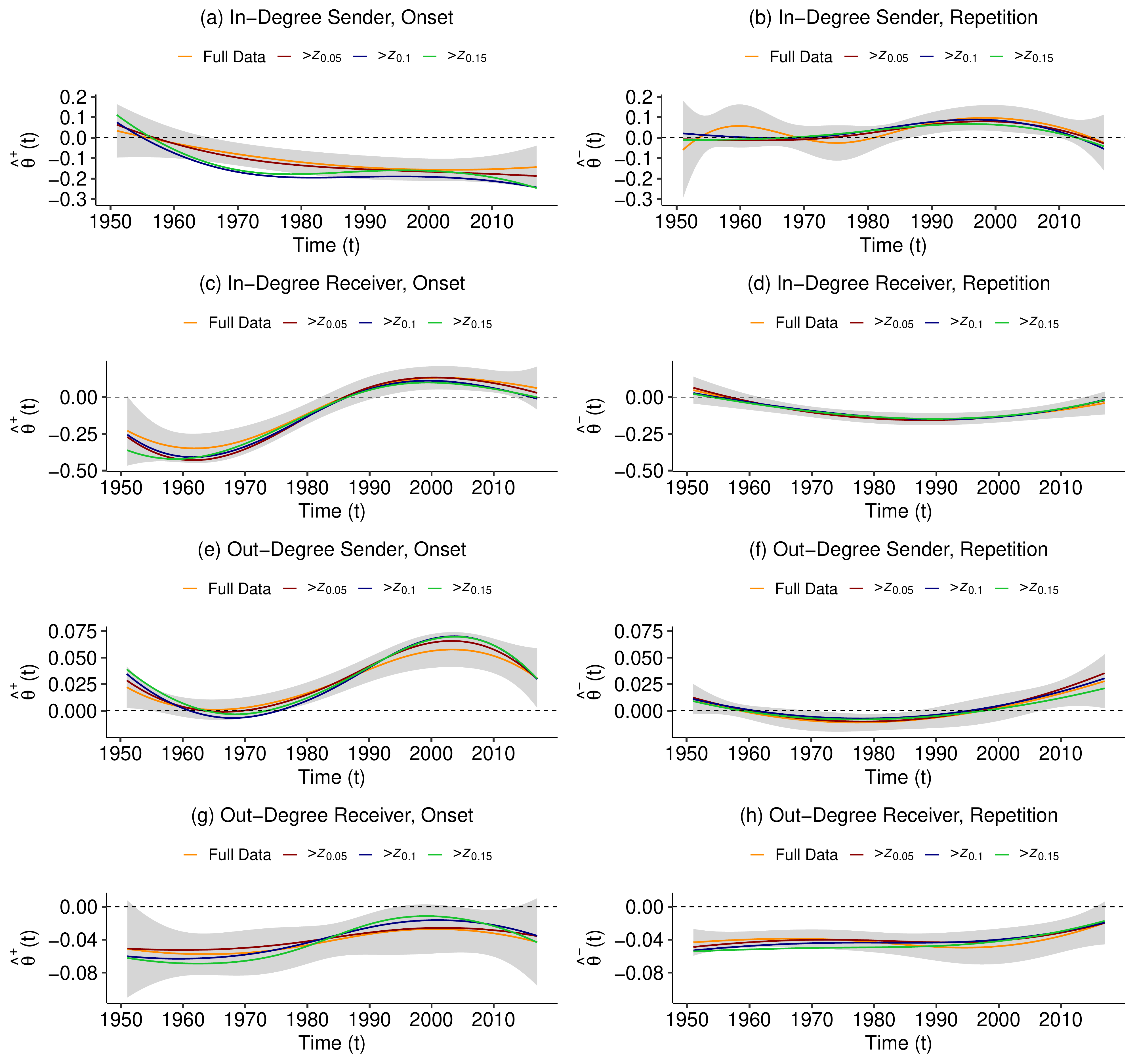}
	\caption{Robustness checks of the estimated parameters when only events with a specific TIV are regarded.The shaded area indicates the $95\%$ confidence bands of the estimates from the unconditional model including all events.}
	\label{fig:comp_end1}
\end{figure}

\begin{figure}[ht!]\centering
	\FloatBarrier
	\includegraphics[trim={0cm 0cm 0cm 0cm},clip,width=\textwidth]{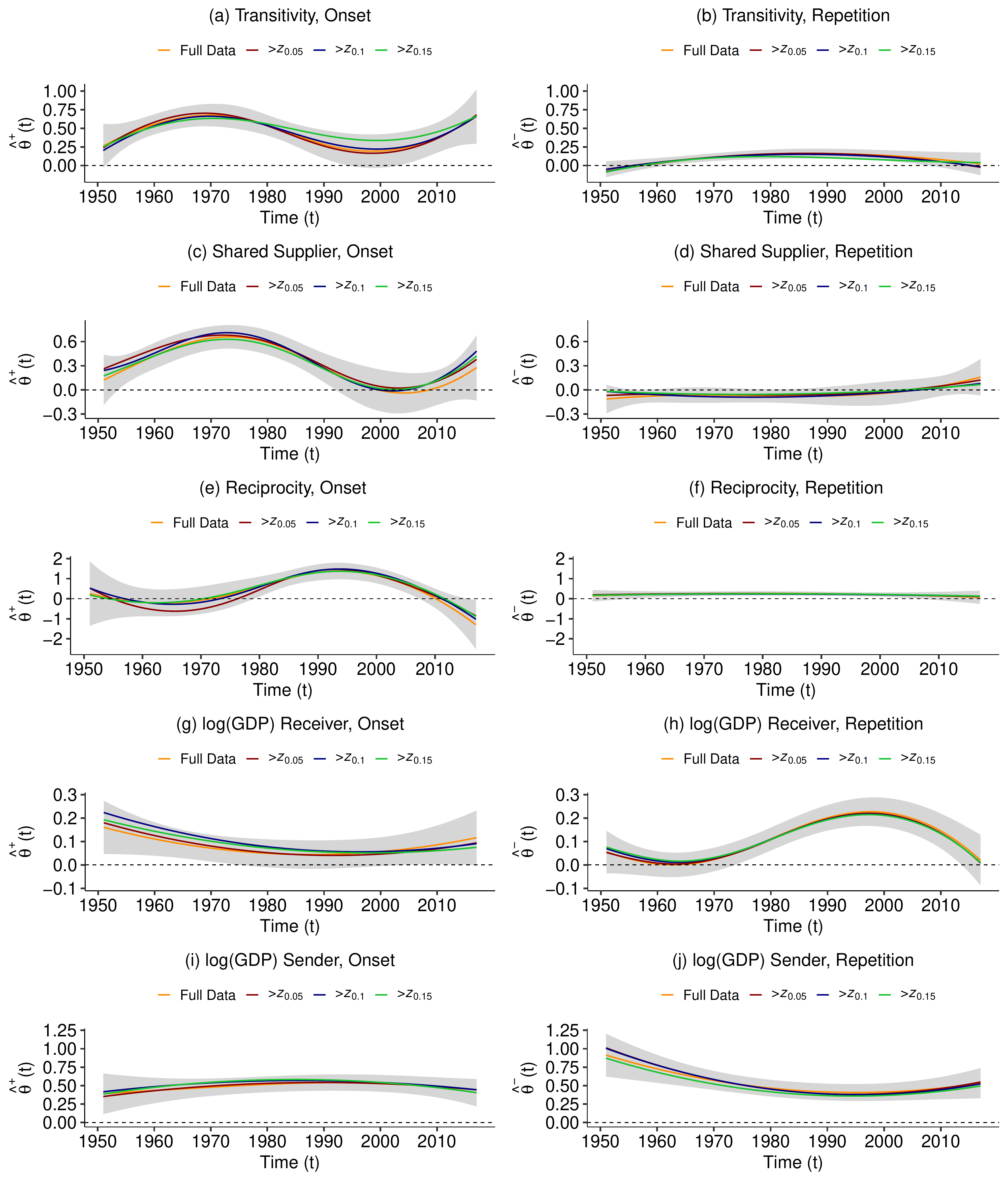}
	\caption{Robustness checks of the estimated parameters when only events with a specific TIV are regarded.The shaded area indicates the $95\%$ confidence bands of the estimates from the unconditional model including all events.}
	\label{fig:comp_end2}
\end{figure}

\begin{figure}[ht!]\centering
	\FloatBarrier
	\includegraphics[trim={0cm 0cm 0cm 0cm},clip,width=\textwidth]{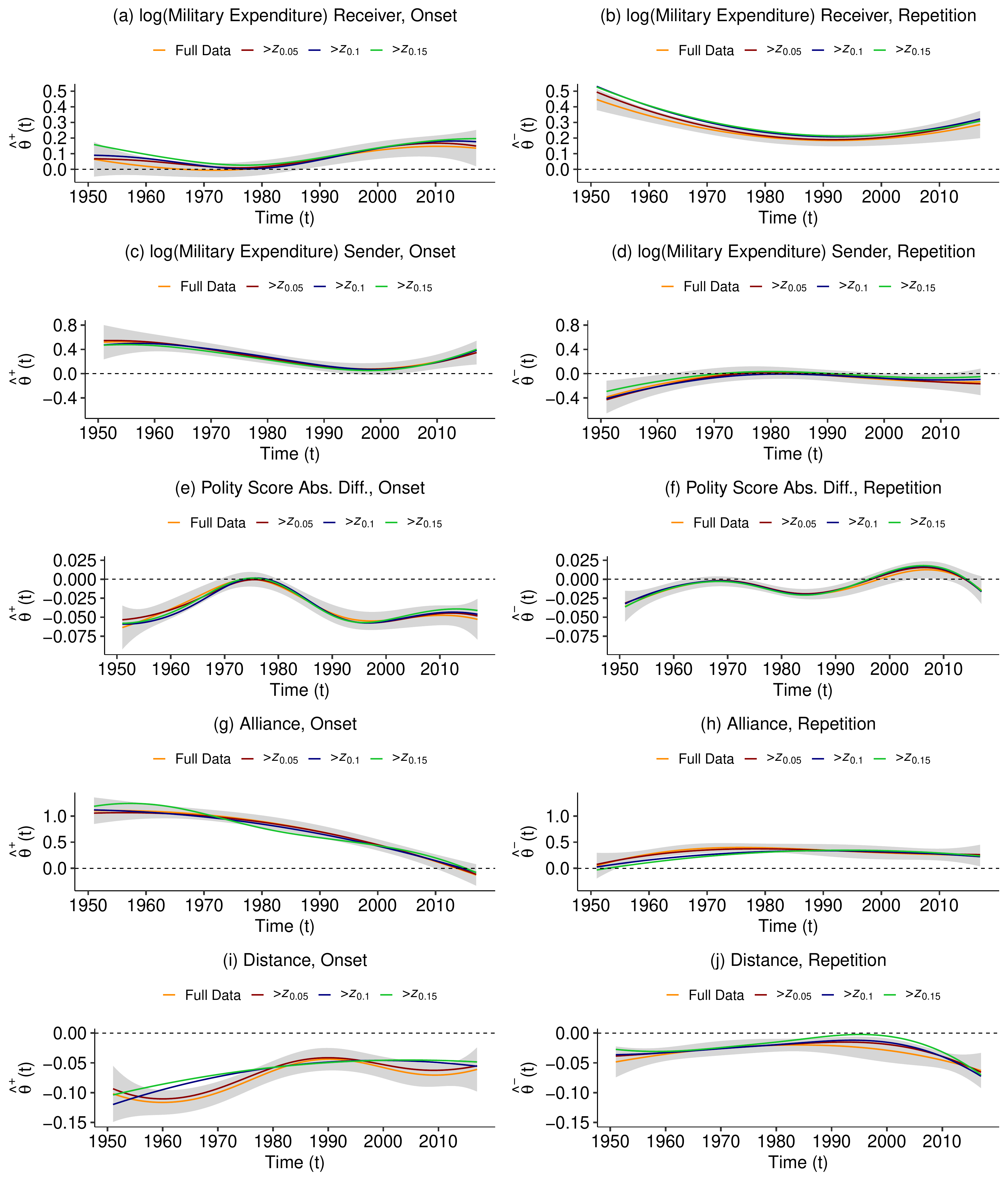}
	\caption{Robustness checks of the estimated parameters when only events with a specific TIV are regarded.The shaded area indicates the $95\%$ confidence bands of the estimates from the unconditional model including all events.}
	\label{fig:comp_exo1}
\end{figure}

The resulting estimates are shown in Figures \ref{fig:comp_end1} to \ref{fig:comp_exo1} and proof the robustness of Figures 4 to 7. More specifically, equal interpretations and conclusions stated in Section 3.3.1 still hold. Only slight variations are visible in Figure \ref{fig:comp_end1} (g) concerning the out-degree of the receiver. Comparing the confidence bands of the original model with the estimates of the conditional models, we observe full coverage in most cases.

	\FloatBarrier

\FloatBarrier

\subsection{Corrected AIC for Finite Sample Size}

	\begin{table}[!]
			\caption{Specifications of the compared models and resulting corrected $\text{AIC}_c$ value.}
		\label{tbl:specs}
	\center
	\begin{tabular}{c c c c c c} 
		& Separability & Time-Varying Effects & Random Effects &   $\text{AIC}_c$    \\ \hline
		Model 1 & \xmark    &       \xmark      &     \xmark     & 84622.47  \\
		Model 2 &  $\checkmark$  &        \xmark        &    \xmark     & 65614.86  \\
		Model 3 &  $\checkmark$  &        $\checkmark$        &   \xmark   & 63174.54 \\
		Model 4 &  $\checkmark$  &      $\checkmark$      &   $\checkmark$   & 59718.04
	\end{tabular} 
\end{table}

Besides correcting for the uncertainty resulting from estimating the variance and tuning parameters of the random and smooth components, we can define a version of the same AIC value that corrects for finite sample sizes as proposed by \citet{hurvich1989}. Table \ref{tbl:specs}  reports this type of AIC value, although the results do not change compared to the values reported in the main article. 

\section{Further Results of the Model Assessment}

	\FloatBarrier
	\begin{figure}[t!]
	\centering
	\includegraphics[width=0.45\linewidth, page = 2]{model_assesment_count.pdf}
	\includegraphics[width=0.45\linewidth, page = 1]{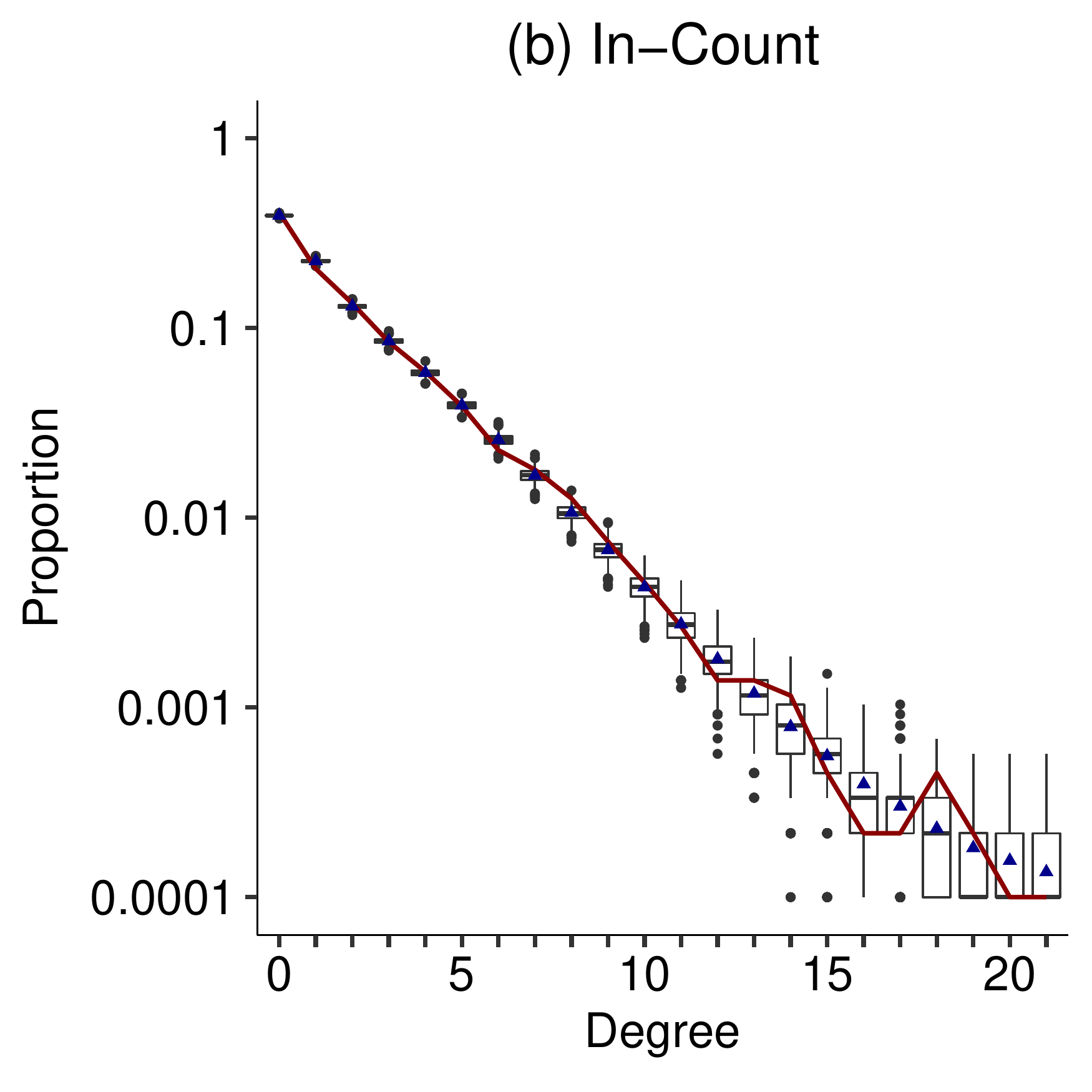}
	\caption{Comparison of the observed and simulated count distributions of the Out- (a) and In-Counts (b) for all included countries summed up over all years. The red lines indicate the observed values of each respective case, whereas the boxplots are the result of drawing 1000 networks and the blue triangles the average values. }
	\label{fig:modelassesmentrootogram}
\end{figure}

We begin by giving the mathematical formulations of the three network statistics for weighted networks analyzed in Section 3.4 of the main article. For the rootogram, we compute the frequencies $h_k$ of combat aircraft deliveries $k \in \lbrace 1, \ldots \rbrace$ over all year.  We calculated the weighted clustering coefficient proposed by \citet{Opsahl2009} for the increments of our network counting process in each year. For the increments $\mathbold y_t$ in year $t$, we hence count the total value of the closed triplets and all triplets and define the generalized clustering coefficient by their ratio. We specify a triplet's value as the arithmetic mean of all observed weights, i.e., the number of yearly deliveries in our application case.  The in-count of all countries in year $t$ determines the yearly average in-count. For country $i$ the in-count in year $t$ is defined by $\text{in-count}(i,t) = \sum_{j = 1}^n y_{ji,t}$. Taking the arithmetic mean over all $\text{in-count}(i,t) ~ \forall~ i\in \mathcal{A}_t$ , where the set $\mathcal{A}_t$ includes all countries present in the trade network in year $t$, gives the average in-count per year. The resultant statistic is proportional to the average events per year. We define the out-count in the same line. If we then concatenate all in- or out-counts over all years, the resulting empirical distribution represents the in- or out-counts irrespective of time. Figure \ref{fig:modelassesmentrootogram} gives visual proof that our model can conserve both the in- and out-count distributions.

\bibliographystyle{apa}
	\bibliography{library}